\documentclass[twocolumn,showpacs,preprintnumbers,prb,amsmath,amssymb]{revtex4}%PRB
\usepackage{graphicx}% Include figure files
\usepackage{dcolumn}% Align table columns on decimal point
\usepackage{bm}% bold math
\usepackage{longtable}% tables which need more space than one page
\usepackage{array}
\usepackage{color}
\begin{document}

% \preprint{}
\title{A scalable qubit architecture based on holes in quantum dot molecules}
\author{Sophia E. Economou$^{1}$}
\author{Juan I. Climente$^{2}$}
\author{Antonio Badolato$^{3}$}
\author{Allan S. Bracker$^{1}$}
\author{Daniel Gammon$^{1}$}
\author{Matthew F. Doty$^{4}$}
\email{doty@udel.edu}

\affiliation{$^{1}$Naval Research Laboratory, Washington, DC 20375}
\affiliation{$^{2}$Dept. de Qu\'imica F\'isica i Anal\'itica, Universitat Jaume I, 12080, Castell\'o, Spain}
\affiliation{$^{3}$Dept. of Physics and Astronomy, University of Rochester, Rochester, NY 14627}
\affiliation{$^{4}$Dept. of Materials Science and Engineering, University of Delaware, Newark, DE 19716}

\date{\today}% It is always \today, today,
             %  but any date may be explicitly specified

\begin{abstract}
Spins confined in quantum dots are a leading candidate for solid
state quantum bits that can be coherently controlled by optical
pulses. There are, however, many challenges to developing a scalable
multi-bit information processing device based on spins in quantum dots,
including the natural inhomogeneous distribution of quantum dot
energy levels, the difficulty of creating all-optical spin
manipulation protocols compatible with nondestructive readout, and
the substantial electron-nuclear hyperfine interaction-induced decoherence. 
Here, we present a scalable qubit design and device architecture based on the spin states of single holes
confined in a quantum dot molecule. The quantum dot molecule qubit
enables a new strategy for optical coherent control with
dramatically enhanced wavelength tunability. The use of hole spins
allows the suppression of decoherence via hyperfine interactions and
enables coherent spin rotations using Raman transitions mediated
by a hole-spin-mixed optically excited state. Because the spin
mixing is present only in the optically excited state, dephasing and
decoherence are strongly suppressed in the ground states that define
the qubits and nondestructive readout is possible. We present the
qubit and device designs and analyze the wavelength tunability and
fidelity of gate operations that can be implemented using this
strategy. We then present experimental and theoretical progress
toward implementing this design.\end{abstract}

% PACS, the Physics and Astronomy
                             % Classification Scheme.
%\keywords{Suggested keywords}%Use showkeys class option if keyword
                              %display desired
\maketitle
% ***************************************************************
% *  Begin Main Text                                            *
% ***************************************************************
Single confined spins have long been considered as possible bit
states for novel optoelectronic logic devices, including quantum
computers.\cite{Imamoglu1999} Spins confined in III-V semiconductor self-assembled quantum dots (QDs) have received a great deal of attention because they interact
strongly with light and provide the opportunity for ultrafast
all-optical implementation of logic operations.\cite{Li2003b,
Chen2004, Economou2006, Economou2007} There has been dramatic progress
in the initialization, coherent manipulation and readout of single
spins in GaAs and InGaAs
QDs,\cite{Atature2006, Berezovsky2008, Press2008, Kim2008, Greilich2009, Kim2010,
Kim2010b, Vamivakas2010} but many challenges to the creation of a
scalable quantum logic device based on optical control of single
spins remain. One serious obstacle is the natural inhomogeneous
distribution of energy levels in a quantum dot ensemble. This
distribution necessitates individually tuned lasers to control each
bit and restricts the ability to mediate entanglement via photonic
cavity modes.\cite{Kim2009} Another obstacle is that existing
approaches to all-optical coherent control of spins
require a transverse magnetic field in order to mix spin states and
permit Raman transitions.\cite{Press2008} The transverse magnetic
field prevents nondestructive readout.\cite{Kim2008} A third
obstacle is the suppression of decoherence. Among the many channels
for decoherence, the hyperfine interaction with nuclei is the most
detrimental.\cite{Fallahi2010, Greilich2011, DeGreve2011}

We present a quantum bit (qubit) design and device architecture that overcomes many of these obstacles. In our approach, the two spin states of a single hole encode the
qubit. The qubit hole is localized in the top QD of a coupled pair of vertically-stacked InAs QDs. These stacked pairs of QDs are known as Quantum Dot Molecules (QDMs) because coherent tunneling leads to the formation of states with delocalized wavefunctions that have molecular symmetries.\cite{Krenner2005, Stinaff2006, Doty2006, Scheibner2007} We propose initializing and rotating the spin using optically excited delocalized molecular states with mixed spin character.  A different optically excited state that is immune to spin mixing provides a recycling transition for readout. We show that this approach enables all-optical coherent control of single hole spins, controlled interactions between two non-identical qubits mediated by photonic cavity modes, and nondestructive readout of the spin projection along the growth axis.

The approach presented here builds on experimental progress in the coherent control of single hole spins\cite{Heiss2007, Gerardot2008, Hsieh2009, Brunner2009, Godden2010, Greilich2011, DeGreve2011, Godden2012} and electron spins in QDMs.\cite{Kim2008} The new design combines advantages of these approaches and makes several key improvements that significantly enhance scalability.  First, the hole spin is less sensitive to dephasing by hyperfine interactions with nuclei, which provide a dominant decoherence mechanism for electrons.\cite{Testelin2009, Greilich2011} Second, the spin mixing for hole tunneling in QDMs can be engineered to be much larger than the mixing for electron tunneling, greatly enhancing the fidelity of the optically-driven qubit rotation. Third, the spin mixing in QDMs provides a pathway to implement all-optical coherent control using only magnetic fields applied along the growth direction, which is parallel to the optical axis. As a result, the approach is compatible with the use of recycling transitions for nondestructive readout. Finally, our approach utilizes optical transitions between electrons and holes located in separate quantum dots. These transitions have an extremely large Stark shift as a function of applied electric fields, and thus provide an enhanced capacity to tune the transitions into resonance with external laser sources or optical cavity modes. 

The use of `indirect' optical transitions, involving electrons and holes in separate QDs, underlies an important conceptual aspect of the device architecture proposed here. Existing device designs often assume an ensemble of identical QDs, which is impossible to achieve experimentally. Demonstrations of spin initialization, control, or readout using QDs typically identify the specific optical transitions of a particular QD and then tune optical cavities and external laser sources into resonance with that particular QD. This approach cannot be scaled beyond a small number of qubits. In our approach we begin by accepting the inhomogeneous distribution of QD energy levels. We design a device architecture in which the optical transitions of a subset of individual QDMs within this ensemble can be tuned into resonance with a fixed-frequency optical cavity or external laser. The key to tuning many individual QDMs into resonance with a fixed-frequency optical cavity or laser is the use of indirect transitions whose energy shifts strongly with the electric field applied locally to individual QDMs. 

We present detailed modeling of the QDM states that could be used for this scheme, compute the fidelity of gate operations implemented with this wavelength-tunable approach and provide experimental evidence of the large spin mixing necessary for such a scheme. In Sect.~\ref{qubit} we present our qubit design and describe how this design suppresses hole spin decoherence
or dephasing in the logical basis states. In Sect.~\ref{strategy} we develop our strategies for full single-qubit all-optical control. We first analyze the use of indirect optical transitions (subsection \ref{indirect}) and then describe how these indirect transitions enable single spin initialization and readout with enhanced wavelength tunability (subsection \ref{initreadout}). In subsection \ref{coherent} we develop the coherent control of the hole spin using Raman transitions mediated by optically excited states with hole spin mixing. In Sect.~\ref{limits} we address scalability in the context of wavelength tunability and fidelity of spin control protocols and discuss how the conditional interactions can be mediated by photonic cavity modes. In Sects.~\ref{ExptEvidence} and \ref{kdotp} we present progress toward the implementation of this device design, including experimental measurements of hole-spin mixing in the optically excited positive trion state and theoretical calculations of QDM structures that enhance the spin mixing.

\section{Qubit Design}\label{qubit}
\begin{figure}[htb]
\begin{center}
\includegraphics[width=8.5cm]{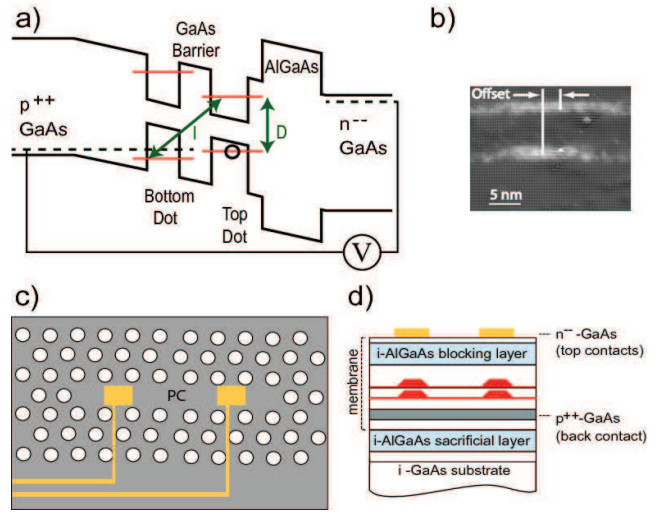}
\caption{(Color Online) a)  Schematic band diagram of proposed qubit
QDM. Direct (D) and Indirect (I) optical transitions are
schematically indicated. b) Cross-sectional STM of a QDM
demonstrating the stacking of QDs and the possibility of lateral
offsets. c) Photonic cavity architecture incorporating multiple
QDMs. d) Schematic cross section of device
heterostructure.\label{QDM_schematic}} \end{center}
\end{figure}

In Fig.~\ref{QDM_schematic}a we schematically depict the band
structure and lowest single particle energy levels for electrons and
holes in the QDM designed for this application. The QDM grown by Molecular Beam Epitaxy is embedded
within a p-i-n diode structure that allows an applied electric field
to tune the energy levels of the QDs relative to one another and the
Fermi level. The bottom QD is
truncated to a smaller height than the top QD so that the hole
levels can be tuned into resonance while electron levels retain a
significant energy offset that causes electrons to always relax to
the top QD.\cite{Bracker2006} The tunnel barrier formed by the intrinsic GaAs region between the p-doped
substrate and the bottom QD is chosen so that the QDM remains
deterministically charged with a single hole in the electric field
range of operation. Tunneling of electrons into the QDM is prevented
by the inclusion of an Al$_x$Ga$_{(1-x)}$As ($x\leq 0.4$) blocking layer. The heterostructure
design is schematically depicted in Fig.~\ref{QDM_schematic}d and
numerical values for the design are discussed in Appendix
\ref{HeterostructureNumbers}.

Our qubit is defined by the spin projections along the growth axis
of a single hole (h$^+$) confined in the QDM. The optically
excited state has one additional electron-hole pair and is called a
positive trion (X$^+$). We use $\left( ^{e_B e_T}_{h_B h_T}\right)$
to describe the spatial location and spin orientation of each
charge: $e_B$ ($e_T$) are the electron spin projections in the
bottom (top) QD; $\uparrow$, $\downarrow$ correspond to $S_e =
\pm1/2$. Holes in QDs contain both light- and heavy-hole contributions and hole spins are properly described as Luttinger spinors.\cite{Climente2008} In single InAs QDs, the heavy-hole-only approximation is largely valid, though the contribution of light holes does impact spin dynamics.\cite{Lu2010} In QDMs the contribution of light-hole states becomes more important and leads to unique and tunable properties for hole spins.\cite{Doty2009, Doty2010, Roloff2010}  The hole spin mixing that emerges in QDMs is a key element of the approach we present here.\cite{Climente2008, Doty2010, Doty2010a} Although the hole spinors contain contributions from all light- and heavy-hole spin projections, the spinors are dominated by a single heavy-hole spin projection. For clarity we label the hole state as $\Uparrow$, $\Downarrow$, corresponding to the dominant heavy hole spin projection ($J_z = \pm3/2$) in each QD. The $\left( ^{e_B e_T}_{h_b h_T}\right)$ notation describes the states of the QDM far away from resonances. When the electric field tunes energy levels into resonance the resulting molecular states can be described as symmetric and antisymmetric combinations of these basis states.

\begin{figure}[htb]
\begin{center}
\includegraphics[width=8.4cm]{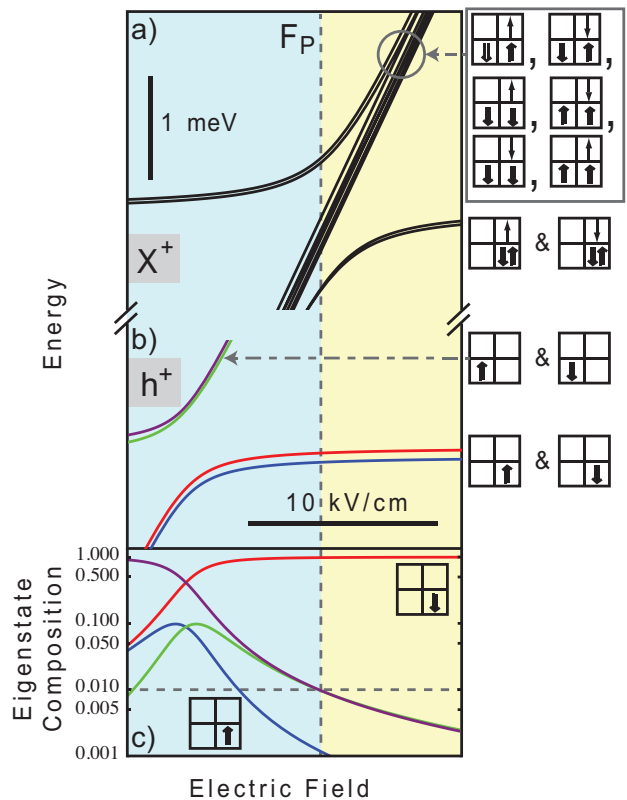}
\caption{(Color Online) a,b)  Calculated energy levels for the $X^+$
(a) and h$^+$ (b) states as a function of applied electric field in
the presence of a 1 T magnetic field parallel to the optical axis. Right column indicates the dominant atomic-like basis states for each group of molecular states. c) Hole spin
contributions to one hole eigenstate as a function of applied electric
field.\label{TrionEnergyLevels}} \end{center}
\end{figure}

In Fig.~\ref{TrionEnergyLevels}a and b we plot the calculated energy
levels of the h$^+$ and X$^+$ states as a function of the applied
electric field in the presence of a static 1 T magnetic field
applied along the growth axis and optical axis (Faraday geometry). The energy level
calculation uses matrix Hamiltonian methods that have been shown to
accurately model the states of QDMs.\cite{Stinaff2006, Doty2006,
Doty2006a, Ponomarev2006, Scheibner2007, Scheibner2007a, Doty2008,
Scheibner2008, Doty2009, Doty2010a} The matrix Hamiltonians and
numerical values used in the calculations presented here are
described in Appendix \ref{ComputationalModel}. We focus first on
the energy levels for the hole (h$^+$). There are four hole states:
$\left(^{0, 0}_{0, \Uparrow}\right)$, $\left(^{0, 0}_{0,
\Downarrow}\right)$, $\left(^{0, 0}_{\Uparrow, 0}\right)$,
$\left(^{0, 0}_{\Downarrow, 0}\right)$. We reference energies to the
energy of a hole in the top QD, so the two states with a hole in
the top QD have an energy that does not depend on electric field
(horizontal lines in Fig.~\ref{TrionEnergyLevels}b). States with a
hole in the bottom QD have a linear dependence of energy on the
applied electric field (diagonal lines in
Fig.~\ref{TrionEnergyLevels}b). When the hole levels are in
resonance, coherent tunneling leads to the formation of molecular
orbitals and the appearance of anticrossings, as shown in
Fig.~\ref{TrionEnergyLevels}b.

The formation of molecular orbitals results in perturbations to the
hole spin g factor and creates the electric field dependent Zeeman
splitting that can be seen in
Fig.~\ref{TrionEnergyLevels}b.\cite{Doty2006} In the absence of hole
spin mixing, the coherent tunneling that results in molecular
orbitals is spin-conserving and the four hole states retain their
spin values throughout the range of electric field. The presence of
a lateral offset between QDs (Fig.~\ref{QDM_schematic}b), however,
breaks the QDM symmetry and creates an effective spin-flip-tunneling
mechanism that mixes hole states with opposite spin projections
located in separate QDs.\cite{Doty2010a} The magnitude of hole spin
mixing included in these calculations (see Appendix
\ref{ComputationalModel}) does not result in the appearance of
anticrossings in Fig.~\ref{TrionEnergyLevels}b because the magnetic
field is too small to bring the spin mixed states sufficiently close
in energy. The presence of the spin mixing, however, appears in
calculations of the hole eigenstate composition. In Fig.~\ref{TrionEnergyLevels}c we
plot the relative contribution of each of the four hole basis states
to the molecular hole state dominated by $\left(^{0, 0}_{0,
\Downarrow}\right)$. Within certain ranges of electric field, the hole spin mixing leads to eigenstates containing significant contributions from multiple spatial and spin configurations. 

We design our qubit and device architecture to take advantage of hole spin mixing in the optically excited state for initialization and control while suppressing the potential negative consequences of hole spin mixing in the logical basis states where information is stored. The logical ground states are the spin projections of a single hole confined in a single QD. The use of holes suppresses hyperfine interactions with nuclei and leads to dephasing times estimated to be at least 100 ns for a hole in a single QD.\cite{Brunner2009} For electric fields to the left of $F_P$ in Fig.~\ref{TrionEnergyLevels}, the logical basis states approach the electric field of coherent tunneling. The formation of molecular states may have negative consequences for the storage of quantum information, including increased spin-orbit interaction, electric-field induced changes in g factor and hole spin mixing.\footnote{We note that all of these effects also provide tools for spin control that may be complementary to the strategy proposed here.} To minimize the potential negative impact of these effects on the storage of quantum information, we design the device architecture to operate only in the range of electric fields to the right of $F_P$ in Fig.~\ref{TrionEnergyLevels} where the logical ground states are localized atomic-like states with minimal perturbations of their spin-orbit interaction, g factor or hole spin mixing.\cite{Climente2008} In Fig.~\ref{TrionEnergyLevels}c we show that the logical basis states remain pure for electric fields to the right of $F_P$: the molecular state is dominated by $\left(^{0,0}_{0, \Downarrow}\right)$ with contributions from  $\left(^{0,
0}_{\Uparrow, 0}\right)$ and $\left(^{0, 0}_{\Downarrow, 0}\right)$ below 1\% and the contribution of $\left(^{0, 0}_{0,\Uparrow}\right)$ below 0.1\%.

The addition of an electron-hole pair to the QDM changes the Coulomb interactions and thus the electric field at which coherent tunneling leads to the formation of molecular states. Our proposed all-optical control strategy operates in the range of electric fields where the logical basis states remain sufficiently pure but the optically excited state can take advantage of hole spin mixing.  As we describe below, hole spin mixing in the optically excited state allows us to perform all of our spin operations without a transverse magnetic field. Consequently, spin projections along the optical axis are eigenstates of the Hamiltonian and our design is compatible with nondestructive readout. In the next sections we develop the strategy for spin initialization, coherent control, and readout.

\section{Full Single qubit Optical Control}\label{strategy}
We propose to manipulate the qubit states with optical pulses that
couple to trion (X$^+$) states. The X$^+$ state contains one
electron located in the top QD and two holes that can be in either
the top or bottom QD. Coulomb interactions with the additional
electron and hole cause the anticrossings for the X$^+$ states to
happen in a different range of electric fields than the
anticrossings of the h$^+$ states, as shown in
Fig.~\ref{TrionEnergyLevels}. There are 12 basis states for the
X$^+$ with unique spatial and spin distributions. The two states
that have two holes in the bottom QD are outside the energy range
we must consider, so only 10 states appear in
Fig.~\ref{TrionEnergyLevels}a, as described in Appendix
\ref{ComputationalModel}. If both holes are located in the top QD,
the Pauli exclusion principle requires them to be in a spin singlet,
e.g. $\left(^{0, \hspace{2pt}\uparrow}_{0,
\Uparrow\Downarrow}\right)$. If the two holes are in separate QDs,
however, both singlet and triplet configurations are
possible.\cite{Stinaff2006, Scheibner2007}  We denote the singlet
state as $\left(^{0,\hspace{1pt}\uparrow}_{\Downarrow,
\Uparrow}\right)_S$ and the three triplet states as
$\left(^{0,\hspace{1pt}\uparrow}_{\Downarrow, \Uparrow}\right)_T$,
$\left(^{0,\hspace{1pt}\uparrow}_{\Uparrow, \Uparrow}\right)$ and
$\left(^{0,\hspace{1pt}\uparrow}_{\Downarrow, \Downarrow}\right)$.
An analogous set of states exist for the electron spin down case
($\downarrow$).

Coherent coupling of any two states leads to anticrossings of the
energy levels. If only spin-conserving tunneling was possible, only
the singlet states would tunnel couple (e.g. $\left(^{0,
\hspace{2pt}\uparrow}_{0, \Uparrow\Downarrow}\right) \leftrightarrow
\left(^{0,\hspace{1pt}\uparrow}_{\Downarrow, \Uparrow}\right)_S$)
and triplet states would pass through the resonance without
coupling. In Sect.~\ref{ExptEvidence} we show experimental evidence
for the existence of hole spin mixing that couples singlet and
triplet states. This hole spin mixing is included in the calculated
energy levels presented in Fig.~\ref{TrionEnergyLevels}. As a result
of this mixing, many of the trion states that appear in
Fig.~\ref{TrionEnergyLevels}a are molecular-like admixtures of
several basis states. We take advantage of this mixing to enable new
optical control strategies.

Figure \ref{StrategyFig} presents a schematic depiction of the
optical transitions used for spin initialization, control, and
readout. These optical transitions are ``indirect'' in that they
couple to X$^+$ states with an electron in the top QD and a hole in
the bottom QD. These indirect transitions are
responsible for the enhanced wavelength tunability we achieve with
this qubit design and control strategy. In Sect.~\ref{indirect} we analyze
the tunability and optical dipole strength of indirect transitions. In Sect.~\ref{initreadout} we analyze
the spin initialization and readout protocols. We develop the
coherent control strategy in Sect.~\ref{coherent}.

\subsection{Indirect transitions}\label{indirect}
\begin{figure}[htb]
\begin{center}
\includegraphics[width=7.0cm]{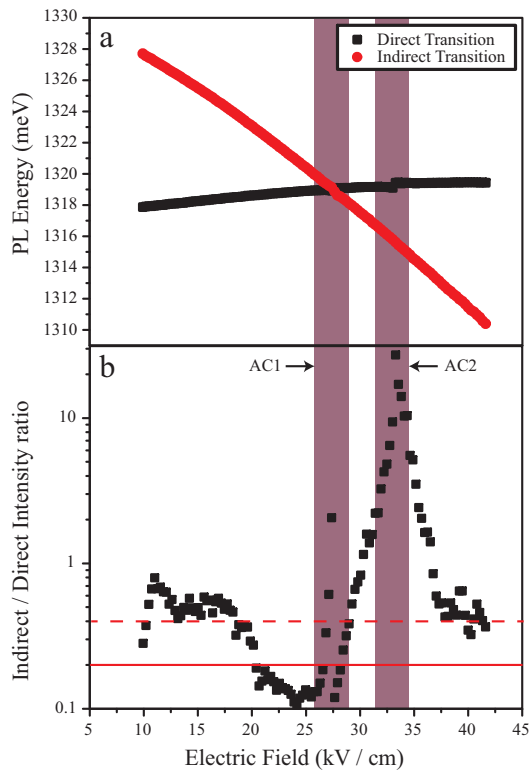}
\caption{(Color Online) a) Energy of direct and indirect PL from the neutral exciton in a single QDM as a function of applied electric field. b) Ratio of PL intensity (indirect / direct) as a function of applied electric field. AC1 indicates the region in which the direct and indirect transitions anticross. AC2 indicates a region in which the energy and intensity of the direct transition is influenced by an anticrossing with an excited indirect transition (not shown).\label{Indirect_graph}} \end{center}
\end{figure}

Direct transitions, as schematically indicated in Fig.~\ref{QDM_schematic}a, involve electrons and holes in the same QDs. These direct transitions (e.g. $\left(^{0,\uparrow}_{0, \Downarrow}\right)$ and $\left(^{0,\downarrow}_{0, \Uparrow}\right)$) have a weak dependence of their energy on the applied electric field due to the quantum confined Stark shift. The energy of the direct transition of the neutral exciton ($X^0$, one electron and one hole) in a QDM with QDs separated by a 4 nm barrier is shown by the black symbols in Fig.~\ref{Indirect_graph}a. Indirect transitions, also depicted in Fig.~\ref{QDM_schematic}a, involve electrons and holes in separate QDs. Because the relative energy levels of the two QDs shift in response to an applied electric field, indirect transitions (e.g. $\left(^{0,\uparrow}_{\Downarrow,0}\right)$ and $\left(^{0,\downarrow}_{\Uparrow,0}\right)$) have a wavelength that depends strongly on the applied electric field. The energy of PL emitted by an indirect $X^0$ transition is shown by the red symbols in Fig.~\ref{Indirect_graph}a. Fig.~\ref{Indirect_graph}a demonstrates that the energy of an indirect transition can be tuned by at least 18 meV as the applied electric field is varied from 5 to 45 kV/cm. 

The sensitivity of indirect transition energies to applied electric field can be enhanced by placing a thicker barrier between the QDs. However, increasing the thickness of the barrier also decreases the optical dipole matrix element of the indirect transition by reducing the overlap between the wavefunctions of the electron and hole. To assess the relative strength of the optical dipole matrix element for indirect transitions, in Fig.~\ref{Indirect_graph}b we plot the ratio of PL intensity emitted by indirect and direct transitions. The intensity of PL emission is not a direct measure of the optical dipole strength because the competing dynamics of radiative recombination and carrier relaxation into and between discrete states influences the probability that charges occupy the different optically excited states. For example, when the indirect transition is at higher energy than the direct transition, we would expect that nonradiative relaxation of the hole (e.g. $\left(^{0,\uparrow}_{\Downarrow,0}\right) \rightarrow \left(^{0,\uparrow}_{0, \Downarrow}\right)$) would favor emission from the direct exciton and reduce the intensity of the indirect transition. In Fig.~\ref{Indirect_graph}b we observe significant fluctuations in the relative intensity of the PL emitted by indirect and direct transitions. The discontinuity near the anticrossing of the direct and indirect transitions (AC1) is partially due to the fact that the states are fully molecular within this range of electric fields and the distinction between direct and indirect does not apply. The fluctuation in the intensity ratio around AC2 arises because the intensity of the direct transition decreases due to an anticrossing with an excited indirect transition (not shown). 

The PL intensity ratio shown in Fig.~\ref{Indirect_graph}b allows us to estimate that the dipole matrix element for indirect transitions in this QDM has an average value approximately 0.4 times the dipole matrix element of the direct transition, as shown by the dashed horizontal line in Fig.~\ref{Indirect_graph}b. We note that there can be significant fluctuations between QDMs in the relative intensity of indirect transitions. The particular QDM studied in Fig.~\ref{Indirect_graph} has a somewhat large indirect transition intensity. This increased intensity for the indirect transition is likely correlated to a weak tunnel coupling in this QDM that inhibits hole relaxation between QDs. This conclusion is supported by the observation of a relatively small anticrossing energy. We take 0.2 as a reasonable approximation of the ratio of dipole matrix elements for indirect and direct optical transitions in QDMs with a 4 nm barrier. This value is depicted by the solid red horizontal line in Fig.~\ref{Indirect_graph}b and is used in our our device design and QDM modeling. Further investigation of the relationship between QDM structure and the dipole matrix element for indirect transitions can guide engineering of QDMs to enhance the coupling of indirect transitions to optical fields. 

\subsection{Initialization and readout} \label{initreadout}
\begin{figure*}[tb]
\begin{center}
\includegraphics[width=18.0cm]{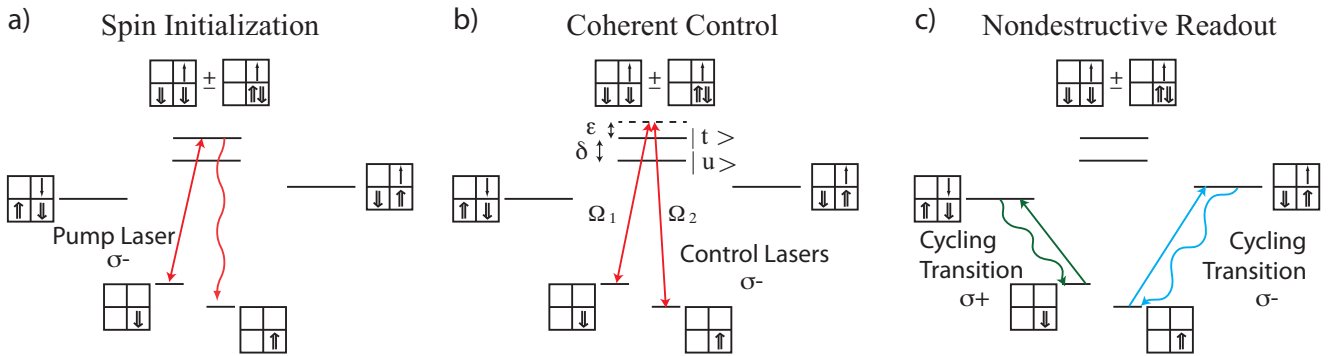}
\caption{(Color Online) Proposed strategy for a) spin initialization,
b) manipulation, and c) readout using optical transitions of
QDMs. Top (bottom) rows indicate the spin projections of electrons
(holes) in the top (right column) and bottom (left column)
QDs.\label{StrategyFig}} \end{center}
\end{figure*}

Our spin initialization protocol (Fig.~\ref{StrategyFig}a) is based
on optical shelving. To initialize our qubit in the hole spin-up
state, we illuminate the QDM with $\sigma^-$ polarized light in
resonance with the optical transition that excites from the
$\left(^{0, 0}_{0, \Downarrow}\right)$ state to one molecular-like
branch of the admixture
$\left(^{0,\hspace{1pt}\uparrow}_{\Downarrow, \Downarrow}\right) \pm
\left(^{0, \hspace{2pt}\uparrow}_{0, \Uparrow\Downarrow}\right)$.
This transition is indicated by the Òpump laserÓ transition in
Fig.~\ref{StrategyFig}a. Because the optically excited state
$\left(^{0,\hspace{1pt}\uparrow}_{\Downarrow, \Downarrow}\right)$ is
indirect, the energy of this optical transition tunes very strongly
with applied electric field, enabling the wavelength tunability we
discuss below. The admixture
$\left(^{0,\hspace{1pt}\uparrow}_{\Downarrow, \Downarrow}\right) \pm
\left(^{0, \hspace{2pt}\uparrow}_{0, \Uparrow\Downarrow}\right)$ is
possible because hole spin mixing allows the spin-flip tunneling
that couples the hole spin down in the bottom QD with the hole spin
up in the top QD. This mixing, and the consequent formation of a
molecular state, is indicated by the +/- joining the two atomic-like
basis states. The relative weight of the two basis states does
depend on the applied electric field, as we discuss further below.

The optically excited state $\left(^{0,\hspace{1pt}\uparrow}_{\Downarrow, \Downarrow}\right) \pm
\left(^{0, \hspace{2pt}\uparrow}_{0, \Uparrow\Downarrow}\right)$ may radiatively decay into the $\left(^{0,0}_{\Downarrow, 0}\right)$ state by emission of a direct exciton. $\left(^{0,0}_{\Downarrow, 0}\right)$ is a metastable state in the range of electric fields considered here because the hole in the ground state of the bottom QD is at higher energy than the ground state of the top QD. The hole will thus relax back to the top QD and be again subject to the spin initialization laser. As a result of the hole spin mixing there is approximately 1\% probability for radiative relaxation from $\left(^{0,\hspace{1pt}\uparrow}_{\Downarrow, \Downarrow}\right) \pm \left(^{0, \hspace{2pt}\uparrow}_{0, \Uparrow\Downarrow}\right)$ to the $\left(^{0, 0}_{0, \Uparrow}\right)$ ground state. After such a radiative decay the hole spin state has been initialized to the spin up projection. The narrow band pump laser is not resonant with any transitions that couple to the hole-spin-up state, so the hole spin is rapidly shelved in the spin-up projection.

Spin readout is achieved by measuring resonance fluorescence as
depicted in Fig.~\ref{StrategyFig}c. The readout states are immune
to hole spin mixing because the Pauli exclusion principle forbids
any state with form $\left(^{0, \hspace{2pt}\downarrow}_{0,
\Uparrow\Uparrow}\right)$. The readout states depicted (e.g.
$\left(^{0,\hspace{1pt}\downarrow}_{\Uparrow, \Downarrow}\right)_S$)
are in fact molecular states (e.g.
$\left(^{0,\hspace{1pt}\downarrow}_{\Uparrow, \Downarrow}\right)_S
\pm \left(^{0, \hspace{2pt}\downarrow}_{0,
\Uparrow\Downarrow}\right)$),  but this spin-conserving mixing does
not alter the polarization selection rules or enable spin flips of
the logical basis states during readout. The direct optical
transitions (electron and hole in same QD) of the readout states are
strongly suppressed because the parallel projections of electron and
hole spin are dark exciton configurations that do not couple to
optical fields. The only mechanism for degradation of the readout
states is a spin-flip of the electron, which can be energetically
suppressed by choosing to readout from the lower energy ($\sigma^+$)
transition. The spin readout transitions are thus Òcycling
transitionsÓ compatible with nondestructive readout: many cycles of
optical absorption and emission can be undertaken to increase
detection probability without risk of altering the spin projection
of the logical basis state.\cite{Kim2008}

\begin{figure}[htb]
\begin{center}
\includegraphics[width=0.8\columnwidth]{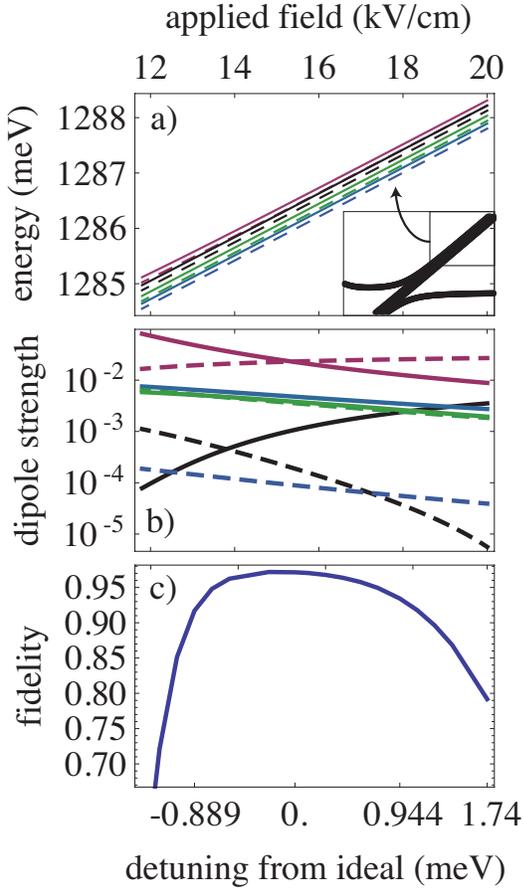}
\caption{(Color Online) a) Energies of optical transitions in our designed QDM as a
function of applied electric field. The two top most lines (red) indicate the transitions
between the two qubit states and the target $X^+$ state
($|t\rangle$). The second highest pair of lines (black) indicate the energy of transitions
involving the unwanted $X^+$ state ($|u\rangle$). Other colors indicate the energies of all other transitions with the same
polarization ($\sigma^-$) that have been included in our calculations. Inset indicates where these transitions lie relative to the anticrossing of the $X^+$ state. b) Dipole strengths of each optical
transition as a function of applied field. Color and dashing as in top panel. Dipole strengths are measured relative to the dipole for a direct transition, which is given a strength of 1.  c) Calculated fidelity of a $\pi/2$ rotation using the two target transitions in the presence of all parasitic transitions and taking into account dipole variations. The lower
axis indicates the QDM's detuning from an ideal laser for each value of the applied electric
field.\label{fidelity}} \end{center}
\end{figure}

Both the spin initialization and spin readout protocols utilize
narrowband lasers resonant with specific optical
transitions. Because the optical transitions used for both
initialization and readout are indirect, the energy of this optical
transition can be varied with the applied electric field to tune the
optical transitions of an arbitrary QDM into resonance with narrow
band initialization and readout lasers utilized for multiple QDMs
within the same device. In Fig.~\ref{fidelity}a we plot the energy
of optical transitions in the QDM designed for this device strategy.
The optical transitions are calculated by taking the difference
between the calculated X$^+$ and h$^+$ state energies plotted in
Fig.~\ref{TrionEnergyLevels}. We plot results in Fig.~\ref{fidelity}
only for electric fields above $F_P$, where the logical basis states
remain suitable for all-optical control.

The results plotted in Fig.~\ref{fidelity}a demonstrate that the
indirect optical transitions can be tuned by 3 meV (from
1285 to 1288 meV) with applied electric fields ranging between
approximately 12 and 20 kV/cm. The linear dependence of wavelength
on applied electric field continues far beyond 20 kV/cm, but we restrict
Fig.~\ref{fidelity} to this range of electric fields for clarity in
the discussion of the coherent control protocol. The maximum
wavelength tunability that can be achieved with indirect transitions
used for spin initialization and readout will likely be limited by
the applied electric field at which the lowest confined hole state
of the bottom QD crosses the first excited hole state of the top QD,
at which point additional spin interactions and/or relaxation
mechanisms could become important. Spectroscopy of the excited
states of holes in QDMs suggests that this limit would permit tuning
over approximately 10 meV.\cite{Scheibner2008}

Tuning over 10 meV is an order of magnitude improvement over the
typical Stark shift tuning (of order 1 meV) achieved for single InAs
QDs\cite{Kim2009} and comparable to the giant Stark shift that can
be achieved for single QDs confined between AlGaAs
barriers.\cite{Bennett2010} In our QDM design the AlGaAs layer that
blocks injection of carriers from the n-type GaAs can be moved
arbitrarily far away from the QDs. As a result, our QDM design is
likely to be more compatible with the fabrication of photonic
crystal cavities than single QDs that achieve a giant Stark shift by
placing AlGaAs layers in close proximity to the QDs. As discussed in Sect.~\ref{indirect}, the dipole
matrix element for indirect transitions is weaker than the comparable dipole matrix element for a direct transition. We discuss the impact of these dipole matrix elements
in more detail below.

\subsection{Spin rotations}\label{coherent}

In this section we develop our approach to the implementation of single qubit
rotations. We quantify the quality of our gates by calculating the
fidelities of our spin rotations in the presence of loss mechanisms
and unwanted dynamics.

To implement arbitrary spin rotations we must be able to perform
rotations about two orthogonal axes; rotations about other axes can
be implemented by combining those. In our present proposed design,
rotations about the quantization ($z$) axis will be carried out using
the cycling (measurement) transitions: a cyclic pulse on the
$\sigma^-$ measurement transition will induce a phase and implement
a rotation about the $z$ axis by an angle determined by the
detuning.\cite{Economou2006,Economou2007} The detuning can be
controlled independently for each qubit state (defined by an individual
QDM) using the electric field that tunes the indirect transition
relative to a fixed optical source frequency. Coherently exciting
the two target (Raman) transitions along the quantization axis with circularly polarized light
allows us to implement rotations about an orthogonal axis.

To implement rotations about the $x$ axis we propose the use of the $\Lambda$
system formed by the two qubit states ($\left(^{0, 0}_{0,
\Downarrow}\right)$ and $\left(^{0, 0}_{0, \Uparrow}\right)$) and
one of the spin-mixed trion states, as depicted in
Fig.~\ref{StrategyFig}b. As shown in Fig.~\ref{StrategyFig}b, there
are two excited states (the spin-mixed trion states
$\alpha_\pm\left(^{0,\hspace{1pt}\uparrow}_{\Downarrow,
\Downarrow}\right) \pm \beta_\pm\left(^{0, \hspace{2pt}\uparrow}_{0,
\Uparrow\Downarrow}\right)$) that are very close in energy. We
develop our strategy based on the use of the high-energy spin-mixed
trion state for the Raman transition and label this state
$|t\rangle$, for target. We want to avoid the low-energy spin-mixed
trion state and label this state $|u\rangle$, for unwanted. We will
return to the impact of this unwanted state below.

The two atomic-like basis states that contribute to the spin-mixed
molecular state $|t\rangle$ involve direct and indirect transitions
with significantly different optical dipoles. A key advantage of our
scheme is that the molecular state that mediates the Raman
transition is dominated by the triplet state that is excited by the
indirect transition ($\left(^{0,\hspace{1pt}\uparrow}_{\Downarrow,
\Downarrow}\right)$). Hole spin mixing adds a small fraction of the
singlet state excited by the direct transition ($\left(^{0,
\hspace{2pt}\uparrow}_{0, \Uparrow\Downarrow}\right)$). Because the
molecular state is dominated by the triplet configuration basis
state, the dipole matrix element for the indirect transition is
comparable to the direct transition. Both transitions have the same
polarization. The energy levels and dipole strengths are computed
assuming a hole spin mixing strength of 300 $\mu$eV, which we
believe to be a value achievable with present growth methods (see
Sect.~\ref{kdotp} and Appendix \ref{ComputationalModel}).

Spin rotations about the $x$ axis are based on coherent population
trapping (CPT). As developed in Ref. [\onlinecite{Economou2007}] we can create an effective two-level
system by using two appropriate phase-locked pulses, each focused on one of the legs
of the $\Lambda$ system. We create an effective two-level system when the parameters of the two pulses are
related in a certain way; \cite{Economou2007} physically, a dark
state is created. When the two transitions have the same Rabi frequency, the
bright state is an equal superposition of $\left(^{0, 0}_{0,
\Downarrow}\right)$ and $\left(^{0, 0}_{0, \Uparrow}\right)$.
Therefore a cyclic evolution will induce a phase to this
superposition state and implement a rotation about the $x$ axis. The angle of rotation can be controlled by the detuning ($\varepsilon$ in Fig.~\ref{StrategyFig}b), which is the same for both legs.

\begin{figure}
\begin{center}
\includegraphics[width=0.98\columnwidth]{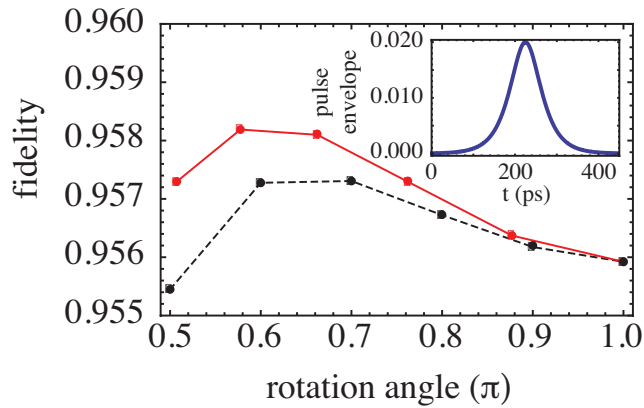}
\caption{Fidelity as function of angle of rotation for `traditional' approach (dashed black line) in which the laser is detuned and our new approach (solid red line) in which the QDM is tuned relative to a fixed laser. The $\pi$-rotation is implemented by a resonant set of pulses, in which case the two approaches coincide. Inset: temporal profile of the hyperbolic secant pulses used in our simulations (intensity in a.u.). The bandwidth of the pulse is $0.02$ meV; the optical decay time is taken to be 1ns.}\label{fidcomp}
\end{center}
\end{figure}

There are always sources of error that will lower the fidelity of the spin
rotations. The finite lifetime of the excited state will introduce
non unitary evolution and therefore will reduce the purity of the
qubit. The unwanted state $|u\rangle$, along with other parasitic
transitions, will effect unitary but unwanted dynamics and will also
lower the quality of the gate. Finally, there is an additional error
coming from the fact that the pulse that is meant to excite one
transition of the $\Lambda$ system will also affect the other one.
This is because the two transitions have the same polarization. To
quantify the effects of these error mechanisms and compare our approach to the `traditional' approach in which the control lasers are tuned relative to the fixed frequency of QD optical transitions, we calculate the
fidelity of the gate. The fidelity is calculated as $\langle Tr(\rho_t\rho)\rangle_\Psi$, where
$\rho_t$ is the density matrix corresponding to the target final
state when starting from initial state $\Psi$, $\rho$ is the actual
final density matrix, and the average is taken over all initial spin
states. The fidelity for the `traditional' approach is plotted by the dashed black line in Fig.~\ref{fidcomp} as a function of the
angle of the $x$-rotation. The angle of rotation is controlled by varying the detuning of the laser while the electric field applied to the QDM, and thus the optical transition energies of the QDM, remain fixed. We choose a fixed electric field of 15.3 kV/cm, where the dipole strengths of the two branches of the Raman transition are equal (see Fig.~\ref{fidelity}b). The inset to Fig.~\ref{fidcomp} shows the temporal pulse envelope used in these calculations. In the next section we assess the impact of changing dipole strengths and electric fields on the fidelity of tunable spin rotations implemented using our approach.

\subsection{Tuning spin rotations for a single QDM}
As discussed above, the angle of coherent spin rotations can be controlled by varying the detuning of the lasers relative to the optical transitions ($\varepsilon$ in Fig.~\ref{StrategyFig}b). In traditional device designs the laser would be tuned relative to the fixed frequency of optical transitions of the QD or QDM that defined the qubit. The fidelity of gate operations that can be achieved for our QDMs using the traditional approach is plotted by the dashed black line in Fig.~\ref{fidcomp}. Our design achieves improved scalability by utilizing laser sources with fixed frequency and tuning individual qubits into resonance with the available sources. The tunability can be used both to select which qubits are affected by the lasers and to control the detuning in order to control the angle of rotation. Unfortunately, the applied electric field that tunes the energies also changes the dipoles of the transitions because the composition of the states varies. As a result, the dipoles do not remain constant when we vary the applied electric field in order to control the detuning. These varying dipoles are shown in Fig.~\ref{fidelity}b. 

To show that the variation in dipole matrix elements does not prohibit execution of spin rotations by arbitrary angles, we calculate the fidelity for spin rotations executed by tuning QDMs relative to a fixed frequency laser source. The computed fidelities are shown by the solid red line in Fig.~\ref{fidcomp}. The calculations include all sources of error considered for the traditional approach in addition to the changing dipole strengths. The laser pulse profile and frequency remain constant and the angle of rotation is varied by tuning the electric field to control the detuning ($\varepsilon$). The plotted value indicates the best fidelity that can be achieved for a given angle of rotation using the design from Ref.~\onlinecite{Economou2007} without additional pulse shaping. To compensate for the unequal dipoles we vary the relative intensity of the two lasers that address the two legs of the $\Lambda$ transition. Rapid modulation of laser power for each target rotation will be challenging, but should be achievable with electrooptic modulators. The optical decay time is taken to be constant at 1 ns because direct radiative recombination will likely remain the dominant lifetime limit. Fig.~\ref{fidcomp} shows that the fidelity of gate operations implemented with our approach is, in fact, slightly better than that achieved by the traditional approach. The electric field must be tuned by only 0.05 kV/cm to affect the change from a $\pi/2$ to a $\pi$ rotation. Fig.~\ref{fidcomp} uses simple pulse shapes and therefore provides a conservative estimate of the fidelities that can be achieved with our approach.

\section{Scalability}\label{limits}
In this section we address two important aspects of scalability. First, we examine the tunability of the
system, in the sense of fixing a modest number of optical sources and tuning the QDMs into resonance with those sources. This approach allows for a substantial decrease in the equipment overhead and device complexity. Second, we develop an approach for coupling arbitrary pairs of qubits without increasing our laser overhead.

\subsection{Tunability in a multi-QDM system}
In the ideal case, all QD-based qubits would be identical and could be addressed by a single group of optical sources. In reality, the inhomogeneous distribution of energy levels in QDs makes this impossible. A mechanism for tuning the optical initialization, control, and readout transitions into resonance with a small number of fixed-frequency optical sources is therefore desirable for the development of a realistic, scalable, multi-qubit device. We have analyzed the wavelength tunability of the spin initialization and readout transitions above, and addressed spin rotation tunability within a single qubit. We now address the wavelength tunability of our spin rotations in the case of a system of inhomogeneous QDMs.

\begin{figure}
\begin{center}
\includegraphics[width=0.98\columnwidth]{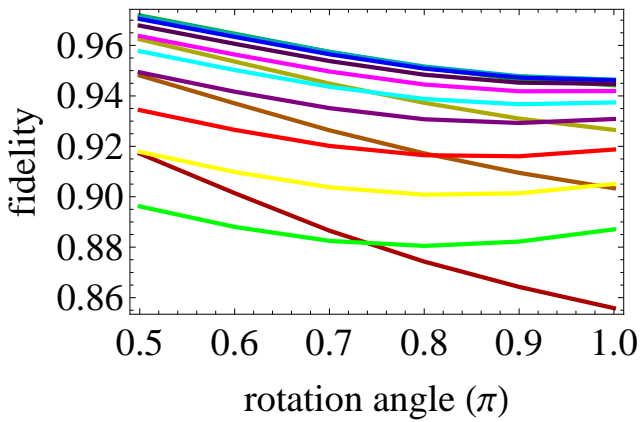}
\caption{(Color online) Fidelity of spin rotation as a function of
the angle of rotation for a multi-QDM system with variations in
their optical resonances. Each curve corresponds to a different
value of the E-field; from top to bottom at $\phi=\pi/2$: 13.8,
14.7, 15.3,  15.6,  16,  16.4, 16.8, 17.2, 13.4, 17.7, 18.1, 19.3 kV/cm.
The pulse envelope is similar to that used in Fig. \ref{fidcomp} but with
half the bandwidth, so about double in temporal
length.}\label{fidtun}
\end{center}
\end{figure}

The inhomogeneous distribution of QD energy levels means that an arbitrary QDM qubit will not have equal dipoles for the two legs of the $\Lambda$ transitions when that QDM qubit is tuned into resonance with the laser sources. As a result, Fig. \ref{fidcomp} is not general enough to describe a multi-QDM system. To analyze the multi-QDM system we compute the fidelity versus rotation angle for various values of the E-field, which simulates the need to tune an individual QDM farther away from the ideal electric field (at which the dipole strengths of the two transitions are equal) in order to tune that QDM into resonance with the available optical source. The results are plotted in Fig.~\ref{fidtun}. The different lines correspond to different applied electric fields with the consequent different relative dipole strengths for the $\Lambda$ transition. In this calculation we do not correlate the detuning of the target transitions to the dipoles. Fig.~\ref{fidtun} therefore describes the range of fidelities that could be achieved for a QDM ensemble with a certain energy distribution. 

To further quantify the range of tuning that could be achieved for an arbitrary QDM qubit, we plot the fidelity for a $\pi/2$ rotation as a function of electric field applied to tune the QDM into resonance with a fixed laser source. The results, presented in Fig.~\ref{fidelity}c, show that our scheme can achieve a fidelity of at least 0.8 when tuning a particular QDM to use a fixed-frequency laser anywhere within a $~$2.5 meV window. A fidelity of 0.8 is certainly well below the formidable values ($\sim$99.99\%) needed for practical quantum computation. The present fidelity values are intended to illustrate that is feasible to tune individual QDMs to utilize fixed-frequency laser sources and should be viewed as a lower bound of what can be achieved. The present calculations have used very simple pulse shapes and there are ways to increase the fidelity of the operations by using numerical pulse optimization techniques to design pulse shapes tailored to this system. In particular, by use of Optimal Control Theory we can impose the condition that the rotation is robust against substantial variations in the ratio of the dipole couplings. This could be done by running a Krotov algorithm in parallel for a number of systems with varying dipoles. Similar approaches can improve fidelities against other types of errors, such as unintended couplings to other states. We expect that with such techniques the fidelities should approach the near-perfect values needed for realistic quantum information processing.

\subsection{Two-qubit gates}
The wavelength tunability enabled by our QDM-based qubit design provides a unique opportunity to couple arbitrary pairs of qubits through a photonic crystal cavity mode. As depicted in Fig.~\ref{QDM_schematic}c and d, the qubits in our proposed device architecture are defined by QDMs that are deterministically spatially-coupled to a single photonic crystal membrane cavity mode.\cite{Badolato2005} The wavelength of the photonic crystal mode can be fine tuned,\cite{Winger2008} but the QD inhomogeneity makes it extremely challenging to choose a cavity resonance that is ideal for all qubits if they are defined by as-grown QDs or QDMs. Our QDM qubit design allows each QDM to be individually controlled by local electric field gates to provide in situ tuning of the coupling between individual QDMs and the cavity mode. We propose to use p- and n-type GaAs as the electrical contacts in order to avoid the inclusion of metals, which would heavily degrade the cavity quality factor.\cite{Imamoglu2007} During device calibration, the local electric field necessary to tune each QDM into resonance with the available optical sources and/or cavity mode would be determined. During the execution of logic operations, each QDM can be tuned appropriately with respect to these optical sources and the cavity.\footnote{We note that conventional photonic cavity designs support linear polarization modes, but circularly polarized modes are advantageous for spin control. Because the cavity is off-resonant from the (narrowband) laser sources, their direct interaction is very weak. The lasers rather interact directly with the QDMs. Therefore, we do not expect the laser polarization to be affected by the cavity.}

Using the approach of Ref.~\onlinecite{Solenov2012}, pairwise entanglement can be created via cavity-mediated entangling gates by selectively tuning two QDMs to the appropriate detuning relative to the cavity mode. The remaining qubits are isolated by significantly detuning them relative to the cavity via use of their local electric fields. The cavity-mediated entanglement, described in more detail below, occurs only when a laser pulse is present, suppressing the possibility of unintended couplings as the QDMs are tuned in preparation for each logical gate operation. The optical transitions relevant to coherent single-spin rotations, readout, and two-qubit gates are all among the highest-energy transitions shown in Fig.~\ref{fidelity}a. Consequently, the QDMs can be red-detuned relative to the cavity when they are to be isolated, ensuring that other optical transitions of the QDM never become more strongly coupled to the cavity mode. 

Our proposed cavity mediated two-qubit gates are based on the readout transitions $\left(^{0, 0}_{0, \Uparrow}\right) \leftrightarrow \left(^{0,\hspace{1pt}\downarrow}_{\Uparrow, \Downarrow}\right)_S \pm \left(^{0, \hspace{2pt}\downarrow}_{0,\Uparrow\Downarrow}\right)$. Bringing these transitions near the two-photon resonance with the cavity,\cite{Solenov2012} we can diagonalize the Hamiltonian consisting of one qubit state and the direct trion excited by a fixed circular polarization for each of two QDMs and by three photonic number states: the zero, one and two photon states. The eigenstates of this Hamiltonian will be entangled states of all three systems (QDM1, QDM2, photon), and will naturally separate into subspaces, depending on the number of excitations.\cite{Solenov2012} These eigenstates have energies that are shifted from the sum of the non-interacting system's net energy, and thus a specific entangled state can be selectively excited optically. The requirement on the cavity is that its $Q$ is high enough that its line broadening is smaller than the Zeeman energy of the qubits and that the QD-cavity coupling $g$ is larger than the decay rates involved. These requirements ensure that the other qubit states will not couple to the cavity and each other and that two-qubit unitary operations are in principle possible. Under these conditions, we can use a single cyclic pulse to induce a $\pi$ phase on the state $|\Uparrow_1\Uparrow_2\rangle$, by which we mean the state of two qubits, each in their own QDM, both with hole spin up. Inducing a $\pi$ phase on the state $|\Uparrow_1\Uparrow_2\rangle$ effects an entangling CZ gate.\cite{Solenov2012}

\section{Experimental evidence of hole-spin mixing in optically excited states of QDMs}\label{ExptEvidence}
As a first step toward implementing this device architecture, we
experimentally measure the magnitude of hole spin mixing in the
positive trion ($X^+$) state. The QDM studied experimentally has a 4
nm barrier separating the two QDs, identical to the barrier in the
QDM proposed here. The QDM we measured is grown on a n-type
substrate and is optically (not electrically) charged with a single
excess hole.\cite{Doty2006, Stinaff2006} The built-in electric field
and QD asymmetry of this sample are also inverted relative to the
sample described above. None of these changes impact the spin mixing
interactions that are the focus of this experiment. The sample is
prepared with an ohmic back contact and Schottky top contacts
enabling us to study individual QDMs using photoluminescence (PL)
spectroscopy as a function of both applied electric and magnetic
fields. A detailed description of experimental procedures can be
found in previous publications.\cite{Doty2010} Here we present the
evidence of hole-spin mixing in the optically excited trion states
($X^+$), which manifests as new anticrossings in the
photoluminescence (PL) spectra. We compare computational simulations
of the observed PL spectra to extract a numerical value of the hole
spin mixing strength.

\begin{figure}[htb]
\begin{center}
\includegraphics[width=8.5cm]{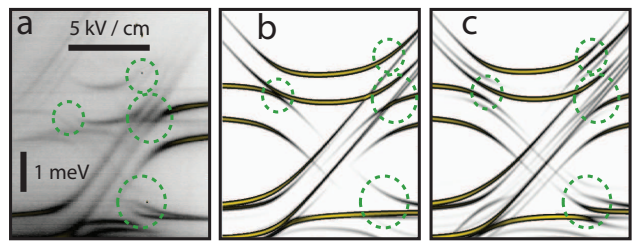}
\caption{(Color Online) Experimental and calculated spectra from the
positive trion in a QDM with a 4 nm barrier in a 6 T longitudinal
magnetic field. a) Experimental spectra, b) Calculation with $hm =
0$, c) Calculation with $hm = 0.2$ meV. The circles highlight four
regions where the features observed experimentally are only
reproduced by the calculation including nonzero
$hm$.\label{postrion}} \end{center}
\end{figure}

In Fig.~\ref{postrion}b we show the experimentally measured PL
spectra of a single QDM in the presence of a 6 T magnetic field. The
figure plots the photoluminescence intensity (grayscale) as a
function of both applied electric field (x axis) and energy (y
axis). The initial and final states of optical recombination are the
$X^+$ and $h^+$ states, analogous to those displayed in
Fig.~\ref{TrionEnergyLevels}a and b. As seen in
Fig.~\ref{TrionEnergyLevels}, Coulomb interactions cause the
anticrossings of the $X^+$ and $h^+$ states to occur at different
values of the applied electric field. As a result, the PL energy as
a function of electric field shows the characteristic ``x" shaped
pattern.\cite{Stinaff2006} Note that the ``x" appears doubled
because the magnetic field introduces a Zeeman splitting between
states that have a total flip of all spins. The circled regions
indicate the locations of anticrossings and fine structure that
arise due to hole spin mixing.

In Figs.~\ref{postrion}b and c we show calculated PL spectra
computed with the matrix Hamiltonian method described in Appendix
\ref{ComputationalModel} and in previous
publications.\cite{Doty2010} In Fig.~\ref{postrion}b the hole spin
mixing term ($hm$) is set to zero and in Fig.~\ref{postrion}c the
hole spin mixing term is set to 0.2 meV. It is clear that the
inclusion of the hole spin mixing term generates the additional
anticrossings and fine structure apparent in the circled regions.
The value of $hm \sim$ 0.2 meV is obtained by varying the magnitude
of $hm$ and evaluating the best fit to the experimental data. This
result experimentally verifies the existence of hole spin mixing in
the trion state and validates the feasibility of the qubit design
and device architecture proposed here.

\section{Controllably generating hole spin mixing with the desired magnitude}\label{kdotp}
For the qubit and device designs we propose, strong spin mixing is
desirable because it separates the target ($|t\rangle$) and unwanted
($|u\rangle$) states. The energy level calculations presented in
Fig.~\ref{TrionEnergyLevels} use a value of $hm=0.3$ meV, and the
energy levels and dipole matrix elements computed with these
parameters are the basis for the calculated fidelity presented in
Fig.~\ref{fidelity}c. The fidelity results would improve for larger
values of $hm$.

As discussed in Ref.~\onlinecite{Doty2010a}, holes are described as
Luttinger spinors with coupled heavy hole (HH) and light hole (LH)
components. The hole pseudospin projections in self-assembled QDs,
$\Uparrow$ and $\Downarrow$, can be identified with the total
angular momentum of the Luttinger spinors, $F_z=3/2$ and $F_z=-3/2$.
When the axial symmetry of the nanostructure is broken, $F_z$ is no
longer a good quantum number and hole spin mixing (i.e. coupling
between $F_z=3/2$ and $F_z=-3/2$ states) arises. The admixing of the
two pseudospins is mediated by the LH components of the
spinor.\cite{Doty2010a} In QDMs, the constituent QDs are usually
symmetric enough to neglect this spin mixing mechanism. If the QDM
is misaligned, however, the application of an electric field forcing
the hole to delocalize over the two QDs leads to a severely
asymmetric orbital and spin mixing becomes important. In
Fig.~\ref{QDM_schematic}b we show a cross-sectional STM of stacked
InAs QDs that shows the possibility of symmetry breaking due to
lateral offset along the stacking axis. Clearly, the larger the
lateral offset between the QDs of the QDM, the stronger the
asymmetry and the stronger the spin mixing. Using k$\cdot$p
calculations we previously showed that a QDM could give rise to
$hm=0.1$ meV for center-to-center lateral offsets of $\sim 5$
nm.\cite{Doty2010a} A 5 nm offset is unusually large but within the
observed range.\cite{Doty2010a}

As described in Sect.~\ref{ExptEvidence}, we have now experimentally
measured values of $hm$ up to $\sim 0.2$ meV. In order to further
increase $hm$ one could use larger offsets, but this structural
parameter is unfortunately difficult to control. Moreover,
increasing the offset has the negative side effect of reducing the
tunneling rate.\cite{Doty2010a} In this section we propose a more
efficient way to enhance the spin mixing by using QDs with larger 
aspect ratio (i.e. larger ratio of height to radius). Larger aspect ratios qualitatively increase the hole
spin mixing because the anisotropic masses of HHs and LHs cause the
LH character to increase with increasing aspect ratio in QDs grown
along the [001] ($z$) direction. As a result of the increased LH character
the pseudospin coupling increases. We analyze the strength of hole
spin mixing as a function of QDM structure using k$\cdot$p theory
and demonstrate that this approach provides a feasible path to
fabrication of QDMs with $hm \geq 0.3$ meV.

The spin mixing parameter $hm$ is defined as the matrix element
coupling states with holes localized in opposite QDs and with
opposite spin. We analyze the dependence of this spin mixing parameter on QDM structure and symmetry with a simplified Hamiltonian. In a basis formed by $|a\rangle=\left(^{0,
0}_{\Uparrow, 0}\right)$ and $|b\rangle=\left(^{0,
0}_{0,\Downarrow}\right)$, the spin mixing Hamiltonian reads:

\begin{equation}
\left(\begin{array}{cc}
E_a + ed\,F & hm\\
hm  & E_b\\
\end{array}\right),\label{1hHamiltonian}
\end{equation}

\noindent where $E_a$ and $E_b$ are the energies of $|a\rangle$ and
$|b\rangle$ at zero electric field. At the resonant electric field
$(E_a+edF)=E_b=E_0$ and the two hybridized states have energies
$E_\pm=E_0 \pm hm$. Thus, $hm=\Delta/2$, where $\Delta=E_+-E_-$ is
the magnitude of the spin anticrossing gap between $\left(^{0,
0}_{\Uparrow, 0}\right)$ and $\left(^{0, 0}_{0,\Downarrow}\right)$.

\begin{figure}[htb]
\begin{center}
\includegraphics[width=8.0cm]{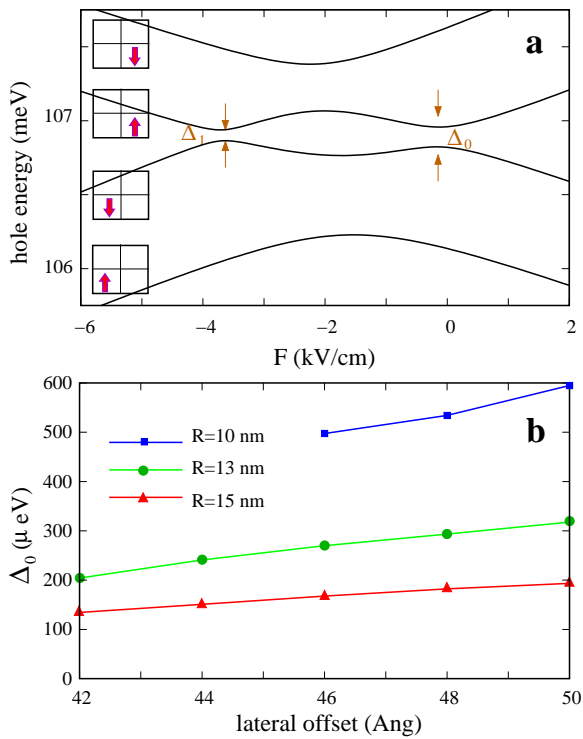}
\caption{(Color Online) (a) Single hole energy as a function of the electric field
in a QDM subject to $B=6$ T.  The QDs have $H=2$ nm, $R=15$ nm,
$D=1.8$ nm and are offset by $4.2$ nm. (b) Spin anticrossing gap as
a function of the QDM offset for QDs with different aspect ratio.
Triangles correspond to $(R,D)=(15,1.8)$ nm, circles to
$(R,D)=(13,1.7)$ nm and squares to $(R,D)=(10,1.6)$
nm.\label{hm_magnitude} } \end{center}
\end{figure}

To determine $\Delta$, we run numerical simulations using the same
theoretical model and material parameters as in
Ref.~\onlinecite{Doty2010a}. A typical hole energy spectrum as a
function of the external electric field is presented in
Fig.~\ref{hm_magnitude}(a). The spectrum is calculated with a
magnetic field $B=6$ T to separate the spin-conserving anticrossings
($F\sim -2$ kV/cm) from the spin anticrossings we are interested in
($F\sim 0$ and $F\sim -4$ kV/cm, see arrows). We note that in our
formulation of the k$\cdot$p Hamiltonian there are no off-diagonal
magnetic terms, so the strength of the HH-LH coupling terms does not
depend on $B$.\cite{Planelles2010}

Fig.~\ref{hm_magnitude}(a) reveals that there are two spin
anticrossings, $\Delta_0$ and $\Delta_1$. In general the magnitude
of the anticrossings is asymmetric, with $\Delta_0 > \Delta_1$. This
is because the spin mixing is mediated by LHs delocalized over the
entire QDM. HHs come into resonance at $F \sim -2$ kV/cm, but LH states come into resonance at electric fields closer to
0 kV/cm because of their lighter vertical mass. Consequently, the $\Delta_0$
anticrossing is closer to the resonant field of LHs, where LH
delocalization and hence spin mixing are maximized.

In Fig.~\ref{hm_magnitude}(b) we plot the magnitude of $\Delta_0$
for different QDMs as a function of the lateral offset. The QDMs are
made of lens-shaped QDs with $H=2$ nm height, radius $R$ and
interdot barrier $D$. Because we are interested in strong spin
mixing, we focus on large (but experimentally accessible) offsets.
For QDs with $R=15$ nm, $\Delta_0$ reaches maximum values of $0.2$
meV ($hm=0.1$ meV), which is the case reported in
Ref.~\onlinecite{Doty2010a}. For QDs with $R=13$ nm $\Delta_0$ is
systematically larger, and $hm=0.15$ meV is attained for a $5$ nm
offset. Last, for QDs with $R=10$ nm the magnitude of the spin
mixing is even stronger and $\Delta_0=0.6$ meV ($hm=0.3$ meV) is
already within reach.

Similar trends to those reported in Fig.~\ref{hm_magnitude}(b) are
obtained for $\Delta_1$, albeit with smaller magnitude ($\Delta_1
\lesssim \Delta_0/2$). One could think of increasing the QD aspect
ratio by growing taller QDs instead of reducing $R$. However, this
also reduces the tunneling rate of the QDM, so reducing the lateral
size is preferable. The lateral size of InAs QDs can be controlled
experimentally by growing on templated surfaces.\cite{Kim2009a} It
may also be possible to use new methods for the growth of QDMs with
precisely tailored three-dimensional configurations to engineer
physical interactions that maximize wavelength tunability and gate
operation fidelity.\cite{Yakes2010}

\section{Summary}
We have presented a scalable qubit design and device
architecture based on the spin states of single holes confined in a
QDM. The QDM qubit enables a new strategy for all-optical coherent
control via indirect optical transitions. The wavelength tunability of indirect transitions allows us to develop a device architecture in which locally-applied
electric fields are used to tune individual QDMs into resonance with
a single photonic cavity mode or a small number of optical sources
at fixed wavelengths. Thus, our design offers a tremendous
reduction in the required overhead as compared to schemes based on
single QDs. We take advantage of hole spin mixing in order to
execute spin initialization and coherent spin rotations with
indirect optical transitions and without transverse magnetic fields.
The absence of transverse magnetic fields enables nondestructive
readout. We show how cycling transitions can be used for
nondestructive readout by optically exciting trion states that are
immune to hole spin mixing. We demonstrate that ultrafast coherent
spin rotations can be achieved by tuning the indirect transitions of individual QDMs into
resonance with a single optical pulse or a pair of pulses. We show that these spin rotations have a minimum fidelity of 0.8 over a minimum tuning range of 2.5 meV, despite fluctuating dipole strengths, and describe a path to the development of optical control protocols that can enhance the fidelity and the wavelength tunability of this approach. We demonstrate experimental and theoretical
progress toward the development of QDM materials to implement the
designs presented and validated here.

\begin{acknowledgments}
SEE acknowledges support from LPS/NSA and in part by ONR, ASB and DG acknowledge
support from NSA/ARO. JIC acknowledges support from MICINN project CTQ2011-27324. MFD acknowledges support from NSF award ECCS-1101754. 

\end{acknowledgments}

\appendix
\section{QDM Design parameters}\label{HeterostructureNumbers}
The substrate doping, choice of QD asymmetry and applied electric
field can be used to control the total charge occupancy and spatial
location of charges in a QDM.\cite{Bracker2006} The qubit we present
here is designed to induce controlled coupling of hole energy levels
in a range of electric fields in which the QDM remains charged with
only one hole. The thickness of the undoped GaAs separating the QDs
from the p-doped region (25 nm) is set by the Coulomb blockade
strength and allows for deterministic charging of the QDM with a
single hole. The QD asymmetry (bottom QD truncated to 2.8 nm, top QD
truncated to 3.2 nm) is chosen to ensure that hole levels come into
resonance for the sign and range of electric fields applied in the
p-i-n diode structure. This asymmetry and sign of the applied
electric field also ensure that the electron is located in the top
QD.

The thickness of the GaAs separating the top QD from the AlGaAs
blocking layer is less critical; it is only necessary to ensure that
that the AlGaAs layer is sufficiently thick to suppress tunneling of
electrons from the n-doped GaAs top contact and sufficiently far away from the QD to avoid altering growth dynamics or hole confinement. Because the thickness
of the GaAs and AlGaAs layers above the top QD are flexible, the net
thickness of the heterostructure can be chosen to provide the
optical confinement necessary to fabricate a high-Q photonic crystal
that places the QDMs at electric field maxima. The Al$_x$Ga$_{(1-x)}$As ($x\geq 0.7$) sacrificial
layer is included below the p-doped GaAs region in order to
facilitate an undercut etch that releases the photonic crystal
membrane.

\section{Computational Model}\label{ComputationalModel}
The energy levels, optical transitions, and optical dipoles
presented above are calculated using matrix Hamiltonian methods
presented in previous publications.\cite{Stinaff2006, Doty2006,
Doty2006a, Ponomarev2006, Scheibner2007, Scheibner2007a, Doty2008,
Scheibner2008, Doty2009, Doty2010a} The relevant basis states for
the optical ground state (logical basis states) are:

\begin{equation}
\begin{array}{cccc}
\left(\begin{array}{cc}
0 & 0 \\
0 & \Uparrow \\
\end{array}\right)

&

\left(\begin{array}{cc}
0 & 0 \\
0 & \Downarrow \\
\end{array}\right)

&

\left(\begin{array}{cc}
0 & 0 \\
\Uparrow & 0\\
\end{array}\right)
&

\left(\begin{array}{cc}
0 & 0 \\
\Downarrow & 0\\
\end{array}\right)
\end{array}
\label{Hole_basis}
\end{equation}

In this basis, the matrix Hamiltonian that describes the hole states is given by:

\begin{widetext}
%\hspace{-1cm}
\begin{equation} \left(\begin{array}{cccc}
 \frac{1}{2} \mu_{B} B g_{h} & 0 & -t_{h} + \frac{1}{2} \mu_{B} B g_{b} & -hm \\
0 & -\frac{1}{2} \mu_{B} B g_{h} & hm & -t_{h} - \frac{1}{2} \mu_{B} B g_{b}\\
-t_{h} + \frac{1}{2} \mu_{B} B g_{b} & hm & dF + \frac{1}{2} \mu_{B} B g_{h} & 0\\
-hm & -t_{h} - \frac{1}{2} \mu_{B} B g_{b} & 0 & dF - \frac{1}{2} \mu_{B} B g_{h}\\
\end{array}\right)\label{HoleHamiltonian}
\end{equation}
\end{widetext}

The definitions and numerical values for each symbol are provided
in Table \ref{numparams}. The basis states for the positive trion
($X^+$) matrix Hamiltonian are:

\begin{equation}
\begin{array}{ccccc}
\left(\begin{array}{cc}
0 & \uparrow \\
0 & \Uparrow \Downarrow\\
\end{array}\right)

&

\left(\begin{array}{cc}
0 & \uparrow\\
\Uparrow & \Downarrow\\
\end{array}\right)_S

&

\left(\begin{array}{cc}
0 & \uparrow\\
\Uparrow & \Downarrow\\
\end{array}\right)_T

&

\left(\begin{array}{cc}
0 & \uparrow\\
\Downarrow & \Downarrow \\
\end{array}\right)

&

\left(\begin{array}{cc}
0 & \uparrow\\
\Uparrow & \Uparrow \\
\end{array}\right)
\end{array}
\label{Trion_basis}
\end{equation}

In this basis, the matrix Hamiltonian that describes the positive trion state is given by:

\begin{widetext}
%\hspace{-1cm}
\begin{equation}E_{X^+} = \left(\begin{array}{ccccc}
\Gamma - \frac{1}{2} \mu_{B} B g_{e}& -t_{X^+} & 0 & hm & hm\\
-t_{X^+} & d F - \frac{1}{2} \mu_{B} B g_{e} & J & 0 & 0\\
0 & J & d F - \frac{1}{2} \mu_{B} B g_{e} & 0 & 0\\
hm & 0 & 0 & d F + J - \frac{1}{2} \mu_{B} B \left(g_{e} + 2 g_{h}\right) & 0\\
hm & 0 & 0 & 0 & d F - J - \frac{1}{2} \mu_{B} B \left(g_{e} - 2 g_{h}\right)\\
\end{array}\right)\label{TrionHamiltonian}
\end{equation}
\end{widetext}

Note that the full matrix is block diagonal with a submatrix for the
electron spin $-1/2$ that exactly parallels the $+1/2$ case
presented here. The parameter definitions and numerical values are
given in Table \ref{numparams}.

%\begin{widetext}
\begin{table}[htb]
\begin{tabular}{|c|c|p{5cm}|}\hline
$E_{X^+}$ & 1280 meV & Energy of the trion\\ \hline
$-t_{h}$ & -0.3 meV & bare hole tunneling strength\\ \hline
$-t_{X^+}$ & -0.45 meV & hole tunneling strength in trion state\\ \hline
$d$ & 4 nm & QD separation\\ \hline
$F$ & varies & applied electric field\\ \hline
$hm$ & 0.3 meV & hole spin mixing strength\\ \hline
$\mu_{B}$ & 57.9 $\mu$eV/T & Bohr magneton\\ \hline
$B$ & 1 T & Magnetic Field\\ \hline
$g_{h}$ & -1.555 & hole g factor\\ \hline
$g_{b}$ & 0.4 & Barrier contribution to hole g factor\\ \hline
$g_{e}$ & -0.64 & electron g factor\\ \hline
$\Gamma$ & 3.2145 meV & Energy shift due to Coulomb interactions when all three charges are in the same QD\\ \hline
$J$ & 0.116 meV & Electron-hole exchange energy\\ \hline
\end{tabular}
\caption{Definitions and numerical values for computational parameters.\label{numparams}}
\end{table}
%\end{widetext}

The energies of hole and trion states are computed by calculating
the eigenvalues of the matrix Hamiltonian as a function of electric
field, $F$. The hole spin purity (Fig.~\ref{TrionEnergyLevels}c) is
calculated by plotting the relative contributions of each basis
state to the eigenvector that describes the admixture of all states
for a given value of $F$. The dipole matrix elements are computed by
the  product $\Psi_{h} \cdot \hat{S} \cdot \Psi_{X^+}$ where
$\Psi_{h}$  and $\Psi_{X^+}$ are the eigenvectors describing the
hole and trion states at a given value of $F$. $\hat{S}$ is a matrix
describing the selection rules for all optical transitions.
$\hat{S}$ assigns amplitude 1 to all bright direct transitions (i.e.
electron and hole in the same QD with opposite spin projections),
amplitude 0 to all dark transitions and amplitude $\frac{1}{2}$ to
all transitions involving singlets or triplets that are
superpositions of a bright and a dark direct transition. Bright
indirect transitions in $\hat{S}$ are given amplitude 0.2,
consistent with the measured ratio of optical intensities for direct
and indirect transitions in samples where the QDs are separated by a
4 nm barrier. Indirect transitions that involve singlets or triplets
that are superpositions of a bright and a dark state are given
amplitude 0.1.

%\bibliography{QuantumDotLibrary}

\begin{thebibliography}{48}
\expandafter\ifx\csname natexlab\endcsname\relax\def\natexlab#1{#1}\fi
\expandafter\ifx\csname bibnamefont\endcsname\relax
  \def\bibnamefont#1{#1}\fi
\expandafter\ifx\csname bibfnamefont\endcsname\relax
  \def\bibfnamefont#1{#1}\fi
\expandafter\ifx\csname citenamefont\endcsname\relax
  \def\citenamefont#1{#1}\fi
\expandafter\ifx\csname url\endcsname\relax
  \def\url#1{\texttt{#1}}\fi
\expandafter\ifx\csname urlprefix\endcsname\relax\def\urlprefix{URL }\fi
\providecommand{\bibinfo}[2]{#2}
\providecommand{\eprint}[2][]{\url{#2}}

\bibitem[{\citenamefont{Imamoglu et~al.}(1999)\citenamefont{Imamoglu,
  Awschalom, Burkard, DiVincenzo, Loss, Sherwin, and Small}}]{Imamoglu1999}
\bibinfo{author}{\bibfnamefont{A.}~\bibnamefont{Imamoglu}},
  \bibinfo{author}{\bibfnamefont{D.~D.} \bibnamefont{Awschalom}},
  \bibinfo{author}{\bibfnamefont{G.}~\bibnamefont{Burkard}},
  \bibinfo{author}{\bibfnamefont{D.~P.} \bibnamefont{DiVincenzo}},
  \bibinfo{author}{\bibfnamefont{D.}~\bibnamefont{Loss}},
  \bibinfo{author}{\bibfnamefont{M.~S.} \bibnamefont{Sherwin}},
  \bibnamefont{and} \bibinfo{author}{\bibfnamefont{A.}~\bibnamefont{Small}},
  \bibinfo{journal}{Physical Review Letters} \textbf{\bibinfo{volume}{83}},
  \bibinfo{pages}{4204} (\bibinfo{year}{1999}).

\bibitem[{\citenamefont{Li et~al.}(2003)\citenamefont{Li, Wu, Steel, Gammon,
  Stievater, Katzer, Park, Piermarocchi, and Sham}}]{Li2003b}
\bibinfo{author}{\bibfnamefont{X.}~\bibnamefont{Li}},
  \bibinfo{author}{\bibfnamefont{Y.}~\bibnamefont{Wu}},
  \bibinfo{author}{\bibfnamefont{D.}~\bibnamefont{Steel}},
  \bibinfo{author}{\bibfnamefont{D.}~\bibnamefont{Gammon}},
  \bibinfo{author}{\bibfnamefont{T.~H.} \bibnamefont{Stievater}},
  \bibinfo{author}{\bibfnamefont{D.~S.} \bibnamefont{Katzer}},
  \bibinfo{author}{\bibfnamefont{D.}~\bibnamefont{Park}},
  \bibinfo{author}{\bibfnamefont{C.}~\bibnamefont{Piermarocchi}},
  \bibnamefont{and} \bibinfo{author}{\bibfnamefont{L.~J.} \bibnamefont{Sham}},
  \bibinfo{journal}{Science} \textbf{\bibinfo{volume}{301}},
  \bibinfo{pages}{809} (\bibinfo{year}{2003}).

\bibitem[{\citenamefont{Chen et~al.}(2004)\citenamefont{Chen, Piermarocchi,
  Sham, Gammon, and Steel}}]{Chen2004}
\bibinfo{author}{\bibfnamefont{P.~C.} \bibnamefont{Chen}},
  \bibinfo{author}{\bibfnamefont{C.}~\bibnamefont{Piermarocchi}},
  \bibinfo{author}{\bibfnamefont{L.~J.} \bibnamefont{Sham}},
  \bibinfo{author}{\bibfnamefont{D.}~\bibnamefont{Gammon}}, \bibnamefont{and}
  \bibinfo{author}{\bibfnamefont{D.~G.} \bibnamefont{Steel}},
  \bibinfo{journal}{Physical Review B} \textbf{\bibinfo{volume}{69}},
  \bibinfo{pages}{75320} (\bibinfo{year}{2004}).

\bibitem[{\citenamefont{Economou et~al.}(2006)\citenamefont{Economou, Sham, Wu,
  and Steel}}]{Economou2006}
\bibinfo{author}{\bibfnamefont{S.}~\bibnamefont{Economou}},
  \bibinfo{author}{\bibfnamefont{L.}~\bibnamefont{Sham}},
  \bibinfo{author}{\bibfnamefont{Y.}~\bibnamefont{Wu}}, \bibnamefont{and}
  \bibinfo{author}{\bibfnamefont{D.}~\bibnamefont{Steel}},
  \bibinfo{journal}{Physical Review B} \textbf{\bibinfo{volume}{74}},
  \bibinfo{pages}{205415} (\bibinfo{year}{2006}).

\bibitem[{\citenamefont{Economou and Reinecke}(2007)}]{Economou2007}
\bibinfo{author}{\bibfnamefont{S.~E.} \bibnamefont{Economou}} \bibnamefont{and}
  \bibinfo{author}{\bibfnamefont{T.~L.} \bibnamefont{Reinecke}},
  \bibinfo{journal}{Physical Review Letters} \textbf{\bibinfo{volume}{99}},
  \bibinfo{pages}{217401} (\bibinfo{year}{2007}).

\bibitem[{\citenamefont{Atat{\"u}re et~al.}(2006)\citenamefont{Atat{\"u}re,
  Dreiser, Badolato, Hogele, Karrai, and Imamoglu}}]{Atature2006}
\bibinfo{author}{\bibfnamefont{M.}~\bibnamefont{Atat{\"u}re}},
  \bibinfo{author}{\bibfnamefont{J.}~\bibnamefont{Dreiser}},
  \bibinfo{author}{\bibfnamefont{A.}~\bibnamefont{Badolato}},
  \bibinfo{author}{\bibfnamefont{A.}~\bibnamefont{Hogele}},
  \bibinfo{author}{\bibfnamefont{K.}~\bibnamefont{Karrai}}, \bibnamefont{and}
  \bibinfo{author}{\bibfnamefont{A.}~\bibnamefont{Imamoglu}},
  \bibinfo{journal}{Science} \textbf{\bibinfo{volume}{312}},
  \bibinfo{pages}{551} (\bibinfo{year}{2006}).

\bibitem[{\citenamefont{Berezovsky et~al.}(2008)\citenamefont{Berezovsky,
  Mikkelsen, Stoltz, Coldren, and Awschalom}}]{Berezovsky2008}
\bibinfo{author}{\bibfnamefont{J.}~\bibnamefont{Berezovsky}},
  \bibinfo{author}{\bibfnamefont{M.}~\bibnamefont{Mikkelsen}},
  \bibinfo{author}{\bibfnamefont{N.}~\bibnamefont{Stoltz}},
  \bibinfo{author}{\bibfnamefont{L.}~\bibnamefont{Coldren}}, \bibnamefont{and}
  \bibinfo{author}{\bibfnamefont{D.}~\bibnamefont{Awschalom}},
  \bibinfo{journal}{Science} \textbf{\bibinfo{volume}{320}},
  \bibinfo{pages}{349} (\bibinfo{year}{2008}).

\bibitem[{\citenamefont{Press et~al.}(2008)\citenamefont{Press, Ladd, Zhang,
  and Yamamoto}}]{Press2008}
\bibinfo{author}{\bibfnamefont{D.}~\bibnamefont{Press}},
  \bibinfo{author}{\bibfnamefont{T.~D.} \bibnamefont{Ladd}},
  \bibinfo{author}{\bibfnamefont{B.}~\bibnamefont{Zhang}}, \bibnamefont{and}
  \bibinfo{author}{\bibfnamefont{Y.}~\bibnamefont{Yamamoto}},
  \bibinfo{journal}{Nature} \textbf{\bibinfo{volume}{456}},
  \bibinfo{pages}{218} (\bibinfo{year}{2008}).

\bibitem[{\citenamefont{Kim et~al.}(2008)\citenamefont{Kim, Economou, Badescu,
  Scheibner, Bracker, Bashkansky, Reinecke, and Gammon}}]{Kim2008}
\bibinfo{author}{\bibfnamefont{D.}~\bibnamefont{Kim}},
  \bibinfo{author}{\bibfnamefont{S.~E.} \bibnamefont{Economou}},
  \bibinfo{author}{\bibfnamefont{S.~C.} \bibnamefont{Badescu}},
  \bibinfo{author}{\bibfnamefont{M.}~\bibnamefont{Scheibner}},
  \bibinfo{author}{\bibfnamefont{A.~S.} \bibnamefont{Bracker}},
  \bibinfo{author}{\bibfnamefont{M.}~\bibnamefont{Bashkansky}},
  \bibinfo{author}{\bibfnamefont{T.~L.} \bibnamefont{Reinecke}},
  \bibnamefont{and} \bibinfo{author}{\bibfnamefont{D.}~\bibnamefont{Gammon}},
  \bibinfo{journal}{Physical Review Letters} \textbf{\bibinfo{volume}{101}},
  \bibinfo{pages}{236804} (\bibinfo{year}{2008}).

\bibitem[{\citenamefont{Greilich et~al.}(2009)\citenamefont{Greilich, Economou,
  Spatzek, Yakovlev, Reuter, Wieck, Reinecke, and Bayer}}]{Greilich2009}
\bibinfo{author}{\bibfnamefont{A.}~\bibnamefont{Greilich}},
  \bibinfo{author}{\bibfnamefont{S.}~\bibnamefont{Economou}},
  \bibinfo{author}{\bibfnamefont{S.}~\bibnamefont{Spatzek}},
  \bibinfo{author}{\bibfnamefont{D.}~\bibnamefont{Yakovlev}},
  \bibinfo{author}{\bibfnamefont{D.}~\bibnamefont{Reuter}},
  \bibinfo{author}{\bibfnamefont{A.}~\bibnamefont{Wieck}},
  \bibinfo{author}{\bibfnamefont{T.}~\bibnamefont{Reinecke}}, \bibnamefont{and}
  \bibinfo{author}{\bibfnamefont{M.}~\bibnamefont{Bayer}},
  \bibinfo{journal}{Nature Physics} \textbf{\bibinfo{volume}{5}},
  \bibinfo{pages}{262} (\bibinfo{year}{2009}).

\bibitem[{\citenamefont{Kim et~al.}(2010)\citenamefont{Kim, Truex, Xu, Sun,
  Steel, Bracker, Gammon, and Sham}}]{Kim2010}
\bibinfo{author}{\bibfnamefont{E.~D.} \bibnamefont{Kim}},
  \bibinfo{author}{\bibfnamefont{K.}~\bibnamefont{Truex}},
  \bibinfo{author}{\bibfnamefont{X.~D.} \bibnamefont{Xu}},
  \bibinfo{author}{\bibfnamefont{B.}~\bibnamefont{Sun}},
  \bibinfo{author}{\bibfnamefont{D.~G.} \bibnamefont{Steel}},
  \bibinfo{author}{\bibfnamefont{A.~S.} \bibnamefont{Bracker}},
  \bibinfo{author}{\bibfnamefont{D.}~\bibnamefont{Gammon}}, \bibnamefont{and}
  \bibinfo{author}{\bibfnamefont{L.~J.} \bibnamefont{Sham}},
  \bibinfo{journal}{Physical Review Letters} \textbf{\bibinfo{volume}{104}},
  \bibinfo{pages}{167401} (\bibinfo{year}{2010}).

\bibitem[{\citenamefont{Kim et~al.}(2011)\citenamefont{Kim, Carter, Greilich,
  Bracker, and Gammon}}]{Kim2010b}
\bibinfo{author}{\bibfnamefont{D.}~\bibnamefont{Kim}},
  \bibinfo{author}{\bibfnamefont{S.}~\bibnamefont{Carter}},
  \bibinfo{author}{\bibfnamefont{A.}~\bibnamefont{Greilich}},
  \bibinfo{author}{\bibfnamefont{A.}~\bibnamefont{Bracker}}, \bibnamefont{and}
  \bibinfo{author}{\bibfnamefont{D.}~\bibnamefont{Gammon}},
  \bibinfo{journal}{Nature Physics} \textbf{\bibinfo{volume}{7}},
  \bibinfo{pages}{223} (\bibinfo{year}{2011}).

\bibitem[{\citenamefont{Vamivakas et~al.}(2010)\citenamefont{Vamivakas, Lu,
  Matthiesen, Zhao, F{\"a}lt, Badolato, and Atat{\"u}re}}]{Vamivakas2010}
\bibinfo{author}{\bibfnamefont{A.}~\bibnamefont{Vamivakas}},
  \bibinfo{author}{\bibfnamefont{C.}~\bibnamefont{Lu}},
  \bibinfo{author}{\bibfnamefont{C.}~\bibnamefont{Matthiesen}},
  \bibinfo{author}{\bibfnamefont{Y.}~\bibnamefont{Zhao}},
  \bibinfo{author}{\bibfnamefont{S.}~\bibnamefont{F{\"a}lt}},
  \bibinfo{author}{\bibfnamefont{A.}~\bibnamefont{Badolato}}, \bibnamefont{and}
  \bibinfo{author}{\bibfnamefont{M.}~\bibnamefont{Atat{\"u}re}},
  \bibinfo{journal}{Nature} \textbf{\bibinfo{volume}{467}},
  \bibinfo{pages}{297} (\bibinfo{year}{2010}).

\bibitem[{\citenamefont{Kim et~al.}(2009{\natexlab{a}})\citenamefont{Kim, Thon,
  Petroff, and Bouwmeester}}]{Kim2009}
\bibinfo{author}{\bibfnamefont{H.}~\bibnamefont{Kim}},
  \bibinfo{author}{\bibfnamefont{S.~M.} \bibnamefont{Thon}},
  \bibinfo{author}{\bibfnamefont{P.~M.} \bibnamefont{Petroff}},
  \bibnamefont{and}
  \bibinfo{author}{\bibfnamefont{D.}~\bibnamefont{Bouwmeester}},
  \bibinfo{journal}{Applied Physics Letters} \textbf{\bibinfo{volume}{95}},
  \bibinfo{pages}{243107} (\bibinfo{year}{2009}{\natexlab{a}}).

\bibitem[{\citenamefont{Fallahi et~al.}(2010)\citenamefont{Fallahi, Yilmaz, and
  Imamoglu}}]{Fallahi2010}
\bibinfo{author}{\bibfnamefont{P.}~\bibnamefont{Fallahi}},
  \bibinfo{author}{\bibfnamefont{S.}~\bibnamefont{Yilmaz}}, \bibnamefont{and}
  \bibinfo{author}{\bibfnamefont{A.}~\bibnamefont{Imamoglu}},
  \bibinfo{journal}{Physical Review Letters} \textbf{\bibinfo{volume}{105}},
  \bibinfo{pages}{257402} (\bibinfo{year}{2010}).

\bibitem[{\citenamefont{Greilich et~al.}(2011)\citenamefont{Greilich, Carter,
  Kim, Bracker, and Gammon}}]{Greilich2011}
\bibinfo{author}{\bibfnamefont{A.}~\bibnamefont{Greilich}},
  \bibinfo{author}{\bibfnamefont{S.~G.} \bibnamefont{Carter}},
  \bibinfo{author}{\bibfnamefont{D.}~\bibnamefont{Kim}},
  \bibinfo{author}{\bibfnamefont{A.~S.} \bibnamefont{Bracker}},
  \bibnamefont{and} \bibinfo{author}{\bibfnamefont{D.}~\bibnamefont{Gammon}},
  \bibinfo{journal}{Nature Photon} \textbf{\bibinfo{volume}{5}},
  \bibinfo{pages}{702} (\bibinfo{year}{2011}).

\bibitem[{\citenamefont{De~Greve et~al.}(2011)\citenamefont{De~Greve, McMahon,
  Press, Ladd, Bisping, Schneider, Kamp, Worschech, Hofling, Forchel
  et~al.}}]{DeGreve2011}
\bibinfo{author}{\bibfnamefont{K.}~\bibnamefont{De~Greve}},
  \bibinfo{author}{\bibfnamefont{P.~L.} \bibnamefont{McMahon}},
  \bibinfo{author}{\bibfnamefont{D.}~\bibnamefont{Press}},
  \bibinfo{author}{\bibfnamefont{T.~D.} \bibnamefont{Ladd}},
  \bibinfo{author}{\bibfnamefont{D.}~\bibnamefont{Bisping}},
  \bibinfo{author}{\bibfnamefont{C.}~\bibnamefont{Schneider}},
  \bibinfo{author}{\bibfnamefont{M.}~\bibnamefont{Kamp}},
  \bibinfo{author}{\bibfnamefont{L.}~\bibnamefont{Worschech}},
  \bibinfo{author}{\bibfnamefont{S.}~\bibnamefont{Hofling}},
  \bibinfo{author}{\bibfnamefont{A.}~\bibnamefont{Forchel}},
  \bibnamefont{et~al.}, \bibinfo{journal}{Nat Phys}
  \textbf{\bibinfo{volume}{7}}, \bibinfo{pages}{872} (\bibinfo{year}{2011}).

\bibitem[{\citenamefont{Krenner et~al.}(2005)\citenamefont{Krenner, Sabathil,
  Clark, Kress, Schuh, Bichler, Abstreiter, and Finley}}]{Krenner2005}
\bibinfo{author}{\bibfnamefont{H.~J.} \bibnamefont{Krenner}},
  \bibinfo{author}{\bibfnamefont{M.}~\bibnamefont{Sabathil}},
  \bibinfo{author}{\bibfnamefont{E.~C.} \bibnamefont{Clark}},
  \bibinfo{author}{\bibfnamefont{A.}~\bibnamefont{Kress}},
  \bibinfo{author}{\bibfnamefont{D.}~\bibnamefont{Schuh}},
  \bibinfo{author}{\bibfnamefont{M.}~\bibnamefont{Bichler}},
  \bibinfo{author}{\bibfnamefont{G.}~\bibnamefont{Abstreiter}},
  \bibnamefont{and} \bibinfo{author}{\bibfnamefont{J.~J.}
  \bibnamefont{Finley}}, \bibinfo{journal}{Physical Review Letters}
  \textbf{\bibinfo{volume}{94}}, \bibinfo{pages}{57402} (\bibinfo{year}{2005}).

\bibitem[{\citenamefont{Stinaff et~al.}(2006)\citenamefont{Stinaff, Scheibner,
  Bracker, Ponomarev, Korenev, Ware, Doty, Reinecke, and Gammon}}]{Stinaff2006}
\bibinfo{author}{\bibfnamefont{E.~A.} \bibnamefont{Stinaff}},
  \bibinfo{author}{\bibfnamefont{M.}~\bibnamefont{Scheibner}},
  \bibinfo{author}{\bibfnamefont{A.~S.} \bibnamefont{Bracker}},
  \bibinfo{author}{\bibfnamefont{I.~V.} \bibnamefont{Ponomarev}},
  \bibinfo{author}{\bibfnamefont{V.~L.} \bibnamefont{Korenev}},
  \bibinfo{author}{\bibfnamefont{M.~E.} \bibnamefont{Ware}},
  \bibinfo{author}{\bibfnamefont{M.~F.} \bibnamefont{Doty}},
  \bibinfo{author}{\bibfnamefont{T.~L.} \bibnamefont{Reinecke}},
  \bibnamefont{and} \bibinfo{author}{\bibfnamefont{D.}~\bibnamefont{Gammon}},
  \bibinfo{journal}{Science} \textbf{\bibinfo{volume}{311}},
  \bibinfo{pages}{636} (\bibinfo{year}{2006}).

\bibitem[{\citenamefont{Doty et~al.}(2006{\natexlab{a}})\citenamefont{Doty,
  Scheibner, Ponomarev, Stinaff, Bracker, Korenev, Reinecke, and
  Gammon}}]{Doty2006}
\bibinfo{author}{\bibfnamefont{M.~F.} \bibnamefont{Doty}},
  \bibinfo{author}{\bibfnamefont{M.}~\bibnamefont{Scheibner}},
  \bibinfo{author}{\bibfnamefont{I.~V.} \bibnamefont{Ponomarev}},
  \bibinfo{author}{\bibfnamefont{E.~A.} \bibnamefont{Stinaff}},
  \bibinfo{author}{\bibfnamefont{A.~S.} \bibnamefont{Bracker}},
  \bibinfo{author}{\bibfnamefont{V.~L.} \bibnamefont{Korenev}},
  \bibinfo{author}{\bibfnamefont{T.~L.} \bibnamefont{Reinecke}},
  \bibnamefont{and} \bibinfo{author}{\bibfnamefont{D.}~\bibnamefont{Gammon}},
  \bibinfo{journal}{Physical Review Letters} \textbf{\bibinfo{volume}{97}},
  \bibinfo{pages}{197202} (\bibinfo{year}{2006}{\natexlab{a}}).

\bibitem[{\citenamefont{Scheibner
  et~al.}(2007{\natexlab{a}})\citenamefont{Scheibner, Doty, Ponomarev, Bracker,
  Stinaff, Korenev, Reinecke, and Gammon}}]{Scheibner2007}
\bibinfo{author}{\bibfnamefont{M.}~\bibnamefont{Scheibner}},
  \bibinfo{author}{\bibfnamefont{M.}~\bibnamefont{Doty}},
  \bibinfo{author}{\bibfnamefont{I.}~\bibnamefont{Ponomarev}},
  \bibinfo{author}{\bibfnamefont{A.}~\bibnamefont{Bracker}},
  \bibinfo{author}{\bibfnamefont{E.}~\bibnamefont{Stinaff}},
  \bibinfo{author}{\bibfnamefont{V.}~\bibnamefont{Korenev}},
  \bibinfo{author}{\bibfnamefont{T.}~\bibnamefont{Reinecke}}, \bibnamefont{and}
  \bibinfo{author}{\bibfnamefont{D.}~\bibnamefont{Gammon}},
  \bibinfo{journal}{Phys. Rev. B} \textbf{\bibinfo{volume}{75}},
  \bibinfo{pages}{245318} (\bibinfo{year}{2007}{\natexlab{a}}).

\bibitem[{\citenamefont{Heiss et~al.}(2007)\citenamefont{Heiss, Schaeck, Huebl,
  Bichler, Abstreiter, Finley, Bulaev, and Loss}}]{Heiss2007}
\bibinfo{author}{\bibfnamefont{D.}~\bibnamefont{Heiss}},
  \bibinfo{author}{\bibfnamefont{S.}~\bibnamefont{Schaeck}},
  \bibinfo{author}{\bibfnamefont{H.}~\bibnamefont{Huebl}},
  \bibinfo{author}{\bibfnamefont{M.}~\bibnamefont{Bichler}},
  \bibinfo{author}{\bibfnamefont{G.}~\bibnamefont{Abstreiter}},
  \bibinfo{author}{\bibfnamefont{J.~J.} \bibnamefont{Finley}},
  \bibinfo{author}{\bibfnamefont{D.~V.} \bibnamefont{Bulaev}},
  \bibnamefont{and} \bibinfo{author}{\bibfnamefont{D.}~\bibnamefont{Loss}},
  \bibinfo{journal}{Physical Review B} \textbf{\bibinfo{volume}{76}},
  \bibinfo{pages}{241306} (\bibinfo{year}{2007}).

\bibitem[{\citenamefont{Gerardot et~al.}(2008)\citenamefont{Gerardot, Brunner,
  Dalgarno, Ohberg, Seidl, Kroner, Karrai, Stoltz, Petroff, and
  Warburton}}]{Gerardot2008}
\bibinfo{author}{\bibfnamefont{B.~D.} \bibnamefont{Gerardot}},
  \bibinfo{author}{\bibfnamefont{D.}~\bibnamefont{Brunner}},
  \bibinfo{author}{\bibfnamefont{P.~A.} \bibnamefont{Dalgarno}},
  \bibinfo{author}{\bibfnamefont{P.}~\bibnamefont{Ohberg}},
  \bibinfo{author}{\bibfnamefont{S.}~\bibnamefont{Seidl}},
  \bibinfo{author}{\bibfnamefont{M.}~\bibnamefont{Kroner}},
  \bibinfo{author}{\bibfnamefont{K.}~\bibnamefont{Karrai}},
  \bibinfo{author}{\bibfnamefont{N.~G.} \bibnamefont{Stoltz}},
  \bibinfo{author}{\bibfnamefont{P.~M.} \bibnamefont{Petroff}},
  \bibnamefont{and} \bibinfo{author}{\bibfnamefont{R.~J.}
  \bibnamefont{Warburton}}, \bibinfo{journal}{Nature}
  \textbf{\bibinfo{volume}{451}}, \bibinfo{pages}{441} (\bibinfo{year}{2008}).

\bibitem[{\citenamefont{Hsieh et~al.}(2009)\citenamefont{Hsieh, Cheriton,
  Korkusinski, and Hawrylak}}]{Hsieh2009}
\bibinfo{author}{\bibfnamefont{C.-Y.} \bibnamefont{Hsieh}},
  \bibinfo{author}{\bibfnamefont{R.}~\bibnamefont{Cheriton}},
  \bibinfo{author}{\bibfnamefont{M.}~\bibnamefont{Korkusinski}},
  \bibnamefont{and} \bibinfo{author}{\bibfnamefont{P.}~\bibnamefont{Hawrylak}},
  \bibinfo{journal}{Physical Review B} \textbf{\bibinfo{volume}{80}},
  \bibinfo{pages}{243107} (\bibinfo{year}{2009}).

\bibitem[{\citenamefont{Brunner et~al.}(2009)\citenamefont{Brunner, Gerardot,
  Dalgarno, Wust, Karrai, Stoltz, Petroff, and Warburton}}]{Brunner2009}
\bibinfo{author}{\bibfnamefont{D.}~\bibnamefont{Brunner}},
  \bibinfo{author}{\bibfnamefont{B.~D.} \bibnamefont{Gerardot}},
  \bibinfo{author}{\bibfnamefont{P.~A.} \bibnamefont{Dalgarno}},
  \bibinfo{author}{\bibfnamefont{G.}~\bibnamefont{Wust}},
  \bibinfo{author}{\bibfnamefont{K.}~\bibnamefont{Karrai}},
  \bibinfo{author}{\bibfnamefont{N.~G.} \bibnamefont{Stoltz}},
  \bibinfo{author}{\bibfnamefont{P.~M.} \bibnamefont{Petroff}},
  \bibnamefont{and} \bibinfo{author}{\bibfnamefont{R.~J.}
  \bibnamefont{Warburton}}, \bibinfo{journal}{Science}
  \textbf{\bibinfo{volume}{325}}, \bibinfo{pages}{70} (\bibinfo{year}{2009}).

\bibitem[{\citenamefont{Godden et~al.}(2010)\citenamefont{Godden, Boyle,
  Ramsay, Fox, and Skolnick}}]{Godden2010}
\bibinfo{author}{\bibfnamefont{T.~M.} \bibnamefont{Godden}},
  \bibinfo{author}{\bibfnamefont{S.~J.} \bibnamefont{Boyle}},
  \bibinfo{author}{\bibfnamefont{A.~J.} \bibnamefont{Ramsay}},
  \bibinfo{author}{\bibfnamefont{A.~M.} \bibnamefont{Fox}}, \bibnamefont{and}
  \bibinfo{author}{\bibfnamefont{M.~S.} \bibnamefont{Skolnick}},
  \bibinfo{journal}{Applied Physics Letters} \textbf{\bibinfo{volume}{97}},
  \bibinfo{pages}{061113} (\bibinfo{year}{2010}).

\bibitem[{\citenamefont{Godden et~al.}(2012)\citenamefont{Godden, Quilter,
  Ramsay, Wu, Brereton, Boyle, Luxmoore, Puebla-Nunez, Fox, and
  Skolnick}}]{Godden2012}
\bibinfo{author}{\bibfnamefont{T.}~\bibnamefont{Godden}},
  \bibinfo{author}{\bibfnamefont{J.}~\bibnamefont{Quilter}},
  \bibinfo{author}{\bibfnamefont{A.}~\bibnamefont{Ramsay}},
  \bibinfo{author}{\bibfnamefont{Y.}~\bibnamefont{Wu}},
  \bibinfo{author}{\bibfnamefont{P.}~\bibnamefont{Brereton}},
  \bibinfo{author}{\bibfnamefont{S.}~\bibnamefont{Boyle}},
  \bibinfo{author}{\bibfnamefont{I.}~\bibnamefont{Luxmoore}},
  \bibinfo{author}{\bibfnamefont{J.}~\bibnamefont{Puebla-Nunez}},
  \bibinfo{author}{\bibfnamefont{A.}~\bibnamefont{Fox}}, \bibnamefont{and}
  \bibinfo{author}{\bibfnamefont{M.}~\bibnamefont{Skolnick}},
  \bibinfo{journal}{Physical Review Letters} \textbf{\bibinfo{volume}{108}},
  \bibinfo{pages}{017402} (\bibinfo{year}{2012}).

\bibitem[{\citenamefont{Testelin et~al.}(2009)\citenamefont{Testelin,
  Bernardot, Eble, and Chamarro}}]{Testelin2009}
\bibinfo{author}{\bibfnamefont{C.}~\bibnamefont{Testelin}},
  \bibinfo{author}{\bibfnamefont{F.}~\bibnamefont{Bernardot}},
  \bibinfo{author}{\bibfnamefont{B.}~\bibnamefont{Eble}}, \bibnamefont{and}
  \bibinfo{author}{\bibfnamefont{M.}~\bibnamefont{Chamarro}},
  \bibinfo{journal}{Physical Review B} \textbf{\bibinfo{volume}{79}},
  \bibinfo{pages}{195440} (\bibinfo{year}{2009}).

\bibitem[{\citenamefont{Bracker et~al.}(2006)\citenamefont{Bracker, Scheibner,
  Doty, Stinaff, Ponomarev, Kim, Whitman, Reinecke, and Gammon}}]{Bracker2006}
\bibinfo{author}{\bibfnamefont{A.~S.} \bibnamefont{Bracker}},
  \bibinfo{author}{\bibfnamefont{M.}~\bibnamefont{Scheibner}},
  \bibinfo{author}{\bibfnamefont{M.~F.} \bibnamefont{Doty}},
  \bibinfo{author}{\bibfnamefont{E.~A.} \bibnamefont{Stinaff}},
  \bibinfo{author}{\bibfnamefont{I.~V.} \bibnamefont{Ponomarev}},
  \bibinfo{author}{\bibfnamefont{J.~C.} \bibnamefont{Kim}},
  \bibinfo{author}{\bibfnamefont{L.~J.} \bibnamefont{Whitman}},
  \bibinfo{author}{\bibfnamefont{T.~L.} \bibnamefont{Reinecke}},
  \bibnamefont{and} \bibinfo{author}{\bibfnamefont{D.}~\bibnamefont{Gammon}},
  \bibinfo{journal}{Applied Physics Letters} \textbf{\bibinfo{volume}{89}},
  \bibinfo{pages}{233110} (\bibinfo{year}{2006}).

\bibitem[{\citenamefont{Climente et~al.}(2008)\citenamefont{Climente,
  Korkusinski, Goldoni, and Hawrylak}}]{Climente2008}
\bibinfo{author}{\bibfnamefont{J.~I.} \bibnamefont{Climente}},
  \bibinfo{author}{\bibfnamefont{M.}~\bibnamefont{Korkusinski}},
  \bibinfo{author}{\bibfnamefont{G.}~\bibnamefont{Goldoni}}, \bibnamefont{and}
  \bibinfo{author}{\bibfnamefont{P.}~\bibnamefont{Hawrylak}},
  \bibinfo{journal}{Physical Review B} \textbf{\bibinfo{volume}{78}},
  \bibinfo{pages}{115323} (\bibinfo{year}{2008}).

\bibitem[{\citenamefont{Lu et~al.}(2010)\citenamefont{Lu, Zhao, Vamivakas,
  Matthiesen, F{\"a}lt, Badolato, and Atat{\"u}re}}]{Lu2010}
\bibinfo{author}{\bibfnamefont{C.~Y.} \bibnamefont{Lu}},
  \bibinfo{author}{\bibfnamefont{Y.}~\bibnamefont{Zhao}},
  \bibinfo{author}{\bibfnamefont{A.~N.} \bibnamefont{Vamivakas}},
  \bibinfo{author}{\bibfnamefont{C.}~\bibnamefont{Matthiesen}},
  \bibinfo{author}{\bibfnamefont{S.}~\bibnamefont{F{\"a}lt}},
  \bibinfo{author}{\bibfnamefont{A.}~\bibnamefont{Badolato}}, \bibnamefont{and}
  \bibinfo{author}{\bibfnamefont{M.}~\bibnamefont{Atat{\"u}re}},
  \bibinfo{journal}{Physical Review B} \textbf{\bibinfo{volume}{81}},
  \bibinfo{pages}{35332} (\bibinfo{year}{2010}).

\bibitem[{\citenamefont{Doty et~al.}(2009)\citenamefont{Doty, Climente,
  Korkusinski, Scheibner, Bracker, Hawrylak, and Gammon}}]{Doty2009}
\bibinfo{author}{\bibfnamefont{M.~F.} \bibnamefont{Doty}},
  \bibinfo{author}{\bibfnamefont{J.~I.} \bibnamefont{Climente}},
  \bibinfo{author}{\bibfnamefont{M.}~\bibnamefont{Korkusinski}},
  \bibinfo{author}{\bibfnamefont{M.}~\bibnamefont{Scheibner}},
  \bibinfo{author}{\bibfnamefont{A.~S.} \bibnamefont{Bracker}},
  \bibinfo{author}{\bibfnamefont{P.}~\bibnamefont{Hawrylak}}, \bibnamefont{and}
  \bibinfo{author}{\bibfnamefont{D.}~\bibnamefont{Gammon}},
  \bibinfo{journal}{Physical Review Letters} \textbf{\bibinfo{volume}{102}},
  \bibinfo{pages}{47401} (\bibinfo{year}{2009}).

\bibitem[{\citenamefont{Doty et~al.}(2010{\natexlab{a}})\citenamefont{Doty,
  Climente, Greilich, Yakes, Bracker, and Gammon}}]{Doty2010}
\bibinfo{author}{\bibfnamefont{M.~F.} \bibnamefont{Doty}},
  \bibinfo{author}{\bibfnamefont{J.~I.} \bibnamefont{Climente}},
  \bibinfo{author}{\bibfnamefont{A.}~\bibnamefont{Greilich}},
  \bibinfo{author}{\bibfnamefont{M.}~\bibnamefont{Yakes}},
  \bibinfo{author}{\bibfnamefont{A.~S.} \bibnamefont{Bracker}},
  \bibnamefont{and} \bibinfo{author}{\bibfnamefont{D.}~\bibnamefont{Gammon}},
  \bibinfo{journal}{Journal of Physics: Conference Series}
  \textbf{\bibinfo{volume}{245}}, \bibinfo{pages}{012002}
  (\bibinfo{year}{2010}{\natexlab{a}}).

\bibitem[{\citenamefont{Roloff et~al.}(2010)\citenamefont{Roloff, Eissfeller,
  Vogl, and P{\"o}tz}}]{Roloff2010}
\bibinfo{author}{\bibfnamefont{R.}~\bibnamefont{Roloff}},
  \bibinfo{author}{\bibfnamefont{T.}~\bibnamefont{Eissfeller}},
  \bibinfo{author}{\bibfnamefont{P.}~\bibnamefont{Vogl}}, \bibnamefont{and}
  \bibinfo{author}{\bibfnamefont{W.}~\bibnamefont{P{\"o}tz}},
  \bibinfo{journal}{New Journal of Physics} \textbf{\bibinfo{volume}{12}},
  \bibinfo{pages}{093012} (\bibinfo{year}{2010}).

\bibitem[{\citenamefont{Doty et~al.}(2010{\natexlab{b}})\citenamefont{Doty,
  Climente, Greilich, Yakes, Bracker, and Gammon}}]{Doty2010a}
\bibinfo{author}{\bibfnamefont{M.~F.} \bibnamefont{Doty}},
  \bibinfo{author}{\bibfnamefont{J.~I.} \bibnamefont{Climente}},
  \bibinfo{author}{\bibfnamefont{A.}~\bibnamefont{Greilich}},
  \bibinfo{author}{\bibfnamefont{M.}~\bibnamefont{Yakes}},
  \bibinfo{author}{\bibfnamefont{A.~S.} \bibnamefont{Bracker}},
  \bibnamefont{and} \bibinfo{author}{\bibfnamefont{D.}~\bibnamefont{Gammon}},
  \bibinfo{journal}{Physical Review B} \textbf{\bibinfo{volume}{81}},
  \bibinfo{pages}{035308} (\bibinfo{year}{2010}{\natexlab{b}}).

\bibitem[{\citenamefont{Doty et~al.}(2006{\natexlab{b}})\citenamefont{Doty,
  Ware, Stinaff, Scheibner, Bracker, Ponomarev, Badescu, Korenev, Reinecke, and
  Gammon}}]{Doty2006a}
\bibinfo{author}{\bibfnamefont{M.~F.} \bibnamefont{Doty}},
  \bibinfo{author}{\bibfnamefont{M.~E.} \bibnamefont{Ware}},
  \bibinfo{author}{\bibfnamefont{E.~A.} \bibnamefont{Stinaff}},
  \bibinfo{author}{\bibfnamefont{M.}~\bibnamefont{Scheibner}},
  \bibinfo{author}{\bibfnamefont{A.~S.} \bibnamefont{Bracker}},
  \bibinfo{author}{\bibfnamefont{I.~V.} \bibnamefont{Ponomarev}},
  \bibinfo{author}{\bibfnamefont{S.~C.} \bibnamefont{Badescu}},
  \bibinfo{author}{\bibfnamefont{V.~L.} \bibnamefont{Korenev}},
  \bibinfo{author}{\bibfnamefont{T.~L.} \bibnamefont{Reinecke}},
  \bibnamefont{and} \bibinfo{author}{\bibfnamefont{D.}~\bibnamefont{Gammon}},
  \bibinfo{journal}{Physica Status Solidi B-Basic Solid State Physics}
  \textbf{\bibinfo{volume}{243}}, \bibinfo{pages}{3859}
  (\bibinfo{year}{2006}{\natexlab{b}}).

\bibitem[{\citenamefont{Ponomarev et~al.}(2006)\citenamefont{Ponomarev,
  Scheibner, Stinaff, Bracker, Doty, Badescu, Ware, Korenev, Reinecke, and
  Gammon}}]{Ponomarev2006}
\bibinfo{author}{\bibfnamefont{I.~V.} \bibnamefont{Ponomarev}},
  \bibinfo{author}{\bibfnamefont{M.}~\bibnamefont{Scheibner}},
  \bibinfo{author}{\bibfnamefont{E.~A.} \bibnamefont{Stinaff}},
  \bibinfo{author}{\bibfnamefont{A.~S.} \bibnamefont{Bracker}},
  \bibinfo{author}{\bibfnamefont{M.~F.} \bibnamefont{Doty}},
  \bibinfo{author}{\bibfnamefont{S.~C.} \bibnamefont{Badescu}},
  \bibinfo{author}{\bibfnamefont{M.~E.} \bibnamefont{Ware}},
  \bibinfo{author}{\bibfnamefont{V.~L.} \bibnamefont{Korenev}},
  \bibinfo{author}{\bibfnamefont{T.~L.} \bibnamefont{Reinecke}},
  \bibnamefont{and} \bibinfo{author}{\bibfnamefont{D.}~\bibnamefont{Gammon}},
  \bibinfo{journal}{Physica Status Solidi B-Basic Solid State Physics}
  \textbf{\bibinfo{volume}{243}}, \bibinfo{pages}{3869} (\bibinfo{year}{2006}).

\bibitem[{\citenamefont{Scheibner
  et~al.}(2007{\natexlab{b}})\citenamefont{Scheibner, Ponomarev, Stinaff, Doty,
  Bracker, Hellberg, Reinecke, and Gammon}}]{Scheibner2007a}
\bibinfo{author}{\bibfnamefont{M.}~\bibnamefont{Scheibner}},
  \bibinfo{author}{\bibfnamefont{I.~V.} \bibnamefont{Ponomarev}},
  \bibinfo{author}{\bibfnamefont{E.~A.} \bibnamefont{Stinaff}},
  \bibinfo{author}{\bibfnamefont{M.~F.} \bibnamefont{Doty}},
  \bibinfo{author}{\bibfnamefont{A.~S.} \bibnamefont{Bracker}},
  \bibinfo{author}{\bibfnamefont{C.~S.} \bibnamefont{Hellberg}},
  \bibinfo{author}{\bibfnamefont{T.~L.} \bibnamefont{Reinecke}},
  \bibnamefont{and} \bibinfo{author}{\bibfnamefont{D.}~\bibnamefont{Gammon}},
  \bibinfo{journal}{Physical Review Letters} \textbf{\bibinfo{volume}{99}},
  \bibinfo{pages}{197402} (\bibinfo{year}{2007}{\natexlab{b}}).

\bibitem[{\citenamefont{Doty et~al.}(2008)\citenamefont{Doty, Scheibner,
  Bracker, and Gammon}}]{Doty2008}
\bibinfo{author}{\bibfnamefont{M.~F.} \bibnamefont{Doty}},
  \bibinfo{author}{\bibfnamefont{M.}~\bibnamefont{Scheibner}},
  \bibinfo{author}{\bibfnamefont{A.~S.} \bibnamefont{Bracker}},
  \bibnamefont{and} \bibinfo{author}{\bibfnamefont{D.}~\bibnamefont{Gammon}},
  \bibinfo{journal}{Physical Review B} \textbf{\bibinfo{volume}{78}},
  \bibinfo{pages}{115316} (\bibinfo{year}{2008}).

\bibitem[{\citenamefont{Scheibner et~al.}(2008)\citenamefont{Scheibner, Yakes,
  Bracker, Ponomarev, Doty, Hellberg, Whitman, Reinecke, and
  Gammon}}]{Scheibner2008}
\bibinfo{author}{\bibfnamefont{M.}~\bibnamefont{Scheibner}},
  \bibinfo{author}{\bibfnamefont{M.}~\bibnamefont{Yakes}},
  \bibinfo{author}{\bibfnamefont{A.~S.} \bibnamefont{Bracker}},
  \bibinfo{author}{\bibfnamefont{I.~V.} \bibnamefont{Ponomarev}},
  \bibinfo{author}{\bibfnamefont{M.~F.} \bibnamefont{Doty}},
  \bibinfo{author}{\bibfnamefont{C.~S.} \bibnamefont{Hellberg}},
  \bibinfo{author}{\bibfnamefont{L.~J.} \bibnamefont{Whitman}},
  \bibinfo{author}{\bibfnamefont{T.~L.} \bibnamefont{Reinecke}},
  \bibnamefont{and} \bibinfo{author}{\bibfnamefont{D.}~\bibnamefont{Gammon}},
  \bibinfo{journal}{Nat Phys} \textbf{\bibinfo{volume}{4}},
  \bibinfo{pages}{291} (\bibinfo{year}{2008}).

\bibitem[{\citenamefont{Bennett et~al.}(2010)\citenamefont{Bennett, Patel,
  Skiba-Szymanska, Nicoll, Farrer, Ritchie, and Shields}}]{Bennett2010}
\bibinfo{author}{\bibfnamefont{A.}~\bibnamefont{Bennett}},
  \bibinfo{author}{\bibfnamefont{R.}~\bibnamefont{Patel}},
  \bibinfo{author}{\bibfnamefont{J.}~\bibnamefont{Skiba-Szymanska}},
  \bibinfo{author}{\bibfnamefont{C.}~\bibnamefont{Nicoll}},
  \bibinfo{author}{\bibfnamefont{I.}~\bibnamefont{Farrer}},
  \bibinfo{author}{\bibfnamefont{D.}~\bibnamefont{Ritchie}}, \bibnamefont{and}
  \bibinfo{author}{\bibfnamefont{A.}~\bibnamefont{Shields}},
  \bibinfo{journal}{Applied Physics Letters} \textbf{\bibinfo{volume}{97}},
  \bibinfo{pages}{031104} (\bibinfo{year}{2010}).

\bibitem[{\citenamefont{Badolato et~al.}(2005)\citenamefont{Badolato, Hennessy,
  Atat{\"u}re, Dreiser, Hu, Petroff, and Imamoglu}}]{Badolato2005}
\bibinfo{author}{\bibfnamefont{A.}~\bibnamefont{Badolato}},
  \bibinfo{author}{\bibfnamefont{K.}~\bibnamefont{Hennessy}},
  \bibinfo{author}{\bibfnamefont{M.}~\bibnamefont{Atat{\"u}re}},
  \bibinfo{author}{\bibfnamefont{J.}~\bibnamefont{Dreiser}},
  \bibinfo{author}{\bibfnamefont{E.}~\bibnamefont{Hu}},
  \bibinfo{author}{\bibfnamefont{P.}~\bibnamefont{Petroff}}, \bibnamefont{and}
  \bibinfo{author}{\bibfnamefont{A.}~\bibnamefont{Imamoglu}},
  \bibinfo{journal}{Science} \textbf{\bibinfo{volume}{308}},
  \bibinfo{pages}{1158} (\bibinfo{year}{2005}).

\bibitem[{\citenamefont{Winger et~al.}(2008)\citenamefont{Winger, Badolato,
  Hennessy, Hu, and Imamogùlu}}]{Winger2008}
\bibinfo{author}{\bibfnamefont{M.}~\bibnamefont{Winger}},
  \bibinfo{author}{\bibfnamefont{A.}~\bibnamefont{Badolato}},
  \bibinfo{author}{\bibfnamefont{K.}~\bibnamefont{Hennessy}},
  \bibinfo{author}{\bibfnamefont{E.}~\bibnamefont{Hu}}, \bibnamefont{and}
  \bibinfo{author}{\bibfnamefont{A.}~\bibnamefont{Imamogùlu}},
  \bibinfo{journal}{Physical review letters} \textbf{\bibinfo{volume}{101}},
  \bibinfo{pages}{226808} (\bibinfo{year}{2008}).

\bibitem[{\citenamefont{Imamoglu et~al.}(2007)\citenamefont{Imamoglu, F{\"a}lt,
  Dreiser, Fernandez, Atat{\"u}re, Hennessy, Badolato, and
  Gerace}}]{Imamoglu2007}
\bibinfo{author}{\bibfnamefont{A.}~\bibnamefont{Imamoglu}},
  \bibinfo{author}{\bibfnamefont{S.}~\bibnamefont{F{\"a}lt}},
  \bibinfo{author}{\bibfnamefont{J.}~\bibnamefont{Dreiser}},
  \bibinfo{author}{\bibfnamefont{G.}~\bibnamefont{Fernandez}},
  \bibinfo{author}{\bibfnamefont{M.}~\bibnamefont{Atat{\"u}re}},
  \bibinfo{author}{\bibfnamefont{K.}~\bibnamefont{Hennessy}},
  \bibinfo{author}{\bibfnamefont{A.}~\bibnamefont{Badolato}}, \bibnamefont{and}
  \bibinfo{author}{\bibfnamefont{D.}~\bibnamefont{Gerace}},
  \bibinfo{journal}{Journal of Applied Physics} \textbf{\bibinfo{volume}{101}},
  \bibinfo{pages}{081602} (\bibinfo{year}{2007}).

\bibitem[{\citenamefont{Solenov et~al.}(2012)\citenamefont{Solenov, Economou,
  and Reinecke}}]{Solenov2012}
\bibinfo{author}{\bibfnamefont{D.}~\bibnamefont{Solenov}},
  \bibinfo{author}{\bibfnamefont{S.~E.} \bibnamefont{Economou}},
  \bibnamefont{and} \bibinfo{author}{\bibfnamefont{T.~L.}
  \bibnamefont{Reinecke}}, \bibinfo{journal}{in review}
  \textbf{\bibinfo{volume}{arXiv:1204.5206v1}} (\bibinfo{year}{2012}).

\bibitem[{\citenamefont{Planelles et~al.}(2010)\citenamefont{Planelles,
  Climente, Rajadell, Doty, Bracker, and Gammon}}]{Planelles2010}
\bibinfo{author}{\bibfnamefont{J.}~\bibnamefont{Planelles}},
  \bibinfo{author}{\bibfnamefont{J.}~\bibnamefont{Climente}},
  \bibinfo{author}{\bibfnamefont{F.}~\bibnamefont{Rajadell}},
  \bibinfo{author}{\bibfnamefont{M.}~\bibnamefont{Doty}},
  \bibinfo{author}{\bibfnamefont{A.}~\bibnamefont{Bracker}}, \bibnamefont{and}
  \bibinfo{author}{\bibfnamefont{D.}~\bibnamefont{Gammon}},
  \bibinfo{journal}{Physical Review B} \textbf{\bibinfo{volume}{82}},
  \bibinfo{pages}{155307} (\bibinfo{year}{2010}).

\bibitem[{\citenamefont{Kim et~al.}(2009{\natexlab{b}})\citenamefont{Kim,
  Sheng, Poole, Dalacu, Lefebvre, Lapointe, Reimer, Aers, and
  Williams}}]{Kim2009a}
\bibinfo{author}{\bibfnamefont{D.}~\bibnamefont{Kim}},
  \bibinfo{author}{\bibfnamefont{W.}~\bibnamefont{Sheng}},
  \bibinfo{author}{\bibfnamefont{P.}~\bibnamefont{Poole}},
  \bibinfo{author}{\bibfnamefont{D.}~\bibnamefont{Dalacu}},
  \bibinfo{author}{\bibfnamefont{J.}~\bibnamefont{Lefebvre}},
  \bibinfo{author}{\bibfnamefont{J.}~\bibnamefont{Lapointe}},
  \bibinfo{author}{\bibfnamefont{M.}~\bibnamefont{Reimer}},
  \bibinfo{author}{\bibfnamefont{G.}~\bibnamefont{Aers}}, \bibnamefont{and}
  \bibinfo{author}{\bibfnamefont{R.}~\bibnamefont{Williams}},
  \bibinfo{journal}{Physical Review B} \textbf{\bibinfo{volume}{79}},
  \bibinfo{pages}{045310} (\bibinfo{year}{2009}{\natexlab{b}}).

\bibitem[{\citenamefont{Yakes et~al.}(2010)\citenamefont{Yakes, Cress,
  Tischler, and Bracker}}]{Yakes2010}
\bibinfo{author}{\bibfnamefont{M.~K.} \bibnamefont{Yakes}},
  \bibinfo{author}{\bibfnamefont{C.~D.} \bibnamefont{Cress}},
  \bibinfo{author}{\bibfnamefont{J.~G.} \bibnamefont{Tischler}},
  \bibnamefont{and} \bibinfo{author}{\bibfnamefont{A.~S.}
  \bibnamefont{Bracker}}, \bibinfo{journal}{ACS Nano}
  \textbf{\bibinfo{volume}{4}}, \bibinfo{pages}{3877} (\bibinfo{year}{2010}).

\end{thebibliography}

\end{document}